\documentclass[lettersize,journal]{IEEEtran}
\usepackage{amsmath,amsfonts}
\usepackage{array}
\usepackage[caption=false,font=normalsize,labelfont=sf,textfont=sf]{subfig}
\usepackage{textcomp}
\usepackage{stfloats}
\usepackage{url}
\usepackage{verbatim}
\usepackage{graphicx}
\usepackage{cite}

\usepackage{algorithm,algorithmicx,algpseudocode}
\usepackage{amssymb}
\usepackage{arydshln}
\usepackage{booktabs}
\usepackage{bm}
\usepackage{cancel}
\usepackage{mathrsfs}
\usepackage{multirow}
\usepackage[dvipsnames]{xcolor}
\usepackage{pifont}
\usepackage{soul}
\sethlcolor{lightgray}

\usepackage{tikz}
\usepackage{pgfplots}
\pgfplotsset{width=7.5cm,compat=1.12}
\usepgfplotslibrary{fillbetween}

\makeatletter
\def\adl@drawiv#1#2#3{%
        \hskip.5\tabcolsep
        \xleaders#3{#2.5\@tempdimb #1{1}#2.5\@tempdimb}%
                #2\z@ plus1fil minus1fil\relax
        \hskip.5\tabcolsep}
\newcommand{\cdashlinelr}[1]{%
  \noalign{\vskip\aboverulesep
           \global\let\@dashdrawstore\adl@draw
           \global\let\adl@draw\adl@drawiv}
  \cdashline{#1}
  \noalign{\global\let\adl@draw\@dashdrawstore
           \vskip\belowrulesep}}
\makeatother

\newcommand{\bgunmask}[1]{\tikz[baseline=(X.base)]{\node(X)[rectangle, fill=white, rounded corners=0.3mm, text height=1.4ex,text depth=-0.5ex]{\textcolor{black}{\hspace{-1pt}#1\hspace{-1pt}}};}}
\newcommand{\bgmask}[1]{\tikz[baseline=(X.base)]{\node(X)[rectangle, fill=gray!10, rounded corners=0.3mm, text height=1.4ex,text depth=-0.5ex]{\textcolor{gray!70}{\hspace{-1pt}#1\hspace{-1pt}}};}}

\DeclareMathOperator*{\argmax}{\text{argmax}}

\newcolumntype{x}[1]{>{\centering\arraybackslash\hspace{0pt}}p{#1}}

\newcommand{\pz}{\phantom{0}}

\definecolor{red}{rgb}{1,0,0}
\newcommand{\red}[1]{{\color{red} #1}}
\definecolor{blue}{rgb}{0,0,1}
\newcommand{\blue}[1]{{\color{blue} #1}}

\begin{document}

\title{End-to-End Speech Recognition with Pre-trained Masked Language Model}

\author{
    Yosuke Higuchi,~\IEEEmembership{Student Member,~IEEE,}
    Tetsuji Ogawa,~\IEEEmembership{Member,~IEEE,}\\
    Tetsunori Kobayashi,~\IEEEmembership{Member,~IEEE,} and
    Shinji Watanabe,~\IEEEmembership{Fellow,~IEEE,}%
    \thanks{Y. Higuchi, T. Ogawa, and T. Kobayashi are with Department of Communications and Computer Engineering, Waseda University, Tokyo 162-0042, Japan (e-mail: \{higuchi, ogawa\}@pcl.cs.waseda.ac.jp, koba@waseda.jp). \par
    S. Watanabe is with Language Technologies Institute, Carnegie Mellon University, Pittsburgh, PA 15213-3891, USA (e-mail: shinjiw@ieee.org).
    }%
}

\maketitle

\begin{abstract}
We present a novel approach to end-to-end automatic speech recognition (ASR) that utilizes pre-trained masked language models (LMs) to
facilitate the extraction of linguistic information.
The proposed models, BERT-CTC and BECTRA,
are specifically designed to effectively integrate pre-trained LMs (e.g., BERT) into end-to-end ASR models.
BERT-CTC adapts BERT for connectionist temporal classification (CTC)
by addressing the constraint of the conditional independence assumption between output tokens.
This enables explicit conditioning of BERT's contextualized embeddings in the ASR process,
seamlessly merging audio and linguistic information through an iterative refinement algorithm.
BECTRA extends BERT-CTC to the transducer framework and
trains the decoder network using a vocabulary suitable for ASR training.
This aims to bridge the gap between the text processed in end-to-end ASR and BERT,
as these models have distinct vocabularies with varying text formats and styles, such as the presence of punctuation.
Experimental results on various ASR tasks demonstrate that the proposed models improve
over both the CTC and transducer-based baselines,
owing to the incorporation of BERT knowledge.
Moreover,
our in-depth analysis and investigation verify the effectiveness of the proposed formulations and architectural designs.
\end{abstract}


\section{Introduction}
\IEEEPARstart{I}{n} the field of natural language processing (NLP),
pre-training of language models (LMs) has emerged as the predominant paradigm.
This approach involves training large-scale LMs on a large quantity of text-only data
using well-designed self-supervised objectives~\cite{devlin2019bert,brown2020language},
thereby enabling the acquisition of versatile linguistic knowledge~\cite{tenney2019bert}.
Such pre-trained models provide sophisticated representations that
enhance the performance of downstream NLP tasks,
while also mitigating the need for extensive supervised training data.
In light of their remarkable success in NLP,
pre-trained LMs have been actively adopted for
a variety of end-to-end speech processing tasks~\cite{shin2019effective,huang2021speech,chuang2020speechbert,chung2021splat,hayashi2019pre,kenter2020improving,bang2022improving},
including end-to-end automatic speech recognition (ASR).

End-to-end ASR aims to model direct speech-to-text conversion
using a single deep neural network (DNN)~\cite{graves2014towards,chorowski2015attention,chan2016listen}.
One of the challenges in end-to-end ASR lies in the notable discrepancy between the characteristics of input and output sequences.
Specifically,
the input sequence is a continuous acoustic signal that contains fine-grained patterns with local dependencies,
while the output sequence comprises discrete linguistic symbols (e.g., subwords or words), exhibiting long-range dependencies.
Such difference in modalities poses a significant effort for end-to-end ASR models
in extracting semantic and morphosyntax information from speech,
which is essential for generating accurate textual output.
Hence, the use of pre-trained LMs holds promising potential
in aiding the extraction of linguistic information for end-to-end ASR.

Several attempts have been made to indirectly employ pre-trained LMs to improve end-to-end ASR models,
such as knowledge distillation~\cite{futami2020distilling,bai2021fast,kubo2022knowledge,lu2022context} and
N-best hypothesis rescoring~\cite{shin2019effective,salazar2020masked,chiu2021innovative,futami2021asr,udagawa2022effect}.
Although these approaches are straightforward and do not interfere with the original end-to-end ASR structures,
they can only benefit from the powerful linguistic knowledge of the LMs
either during training or inference.
More recently, there have been efforts to integrate pre-trained LMs directly into end-to-end ASR models,
accomplished by fine-tuning the LMs in conjunction with a speech processing network~\cite{huang2021speech,yi2021efficiently,zheng2021wav,deng2021improving,yu2022non}.
This enables explicit adaptation of pre-trained LMs to ASR,
while allowing models to exploit the linguistic knowledge during both training and inference.
However, these approaches require a complex mechanism to summarize the speech input
into a sequence of appropriate output length before it can be fed into the LMs.
Moreover, to effectively optimize the unified model,
the fine-tuning process entails precise calibration and scheduling of hyperparameters.

In this paper,
we present a novel approach for integrating a pre-trained masked LM (e.g., BERT~\cite{devlin2019bert}) into end-to-end ASR.
To achieve this, we propose two models, \textbf{BERT-CTC} and \textbf{BECTRA},
specifically designed to overcome the challenges associated with the integration.
BERT-CTC facilitates the combination of audio and linguistic features,
based on the formulation of connectionist temporal classification (CTC)~\cite{graves2006connectionist}.
More precisely, BERT embeddings are used to explicitly condition CTC on context-aware linguistic information,
thereby mitigating the conditional independence in outputs.
BERT-CTC exploits the capabilities of BERT without requiring fine-tuning,
while enabling end-to-end training and inference using BERT knowledge and
retaining the advantages of the efficient CTC framework.
BERT-CTC-Transducer (BECTRA) is an extension of BERT-CTC developed to address the discrepancy
between text formats and styles employed in end-to-end ASR and BERT.
BECTRA expands BERT-CTC to the transducer-based model~\cite{graves2012sequence} and
trains the decoder (i.e., prediction/joint networks) using a vocabulary tailored to the target ASR task.
This distinct decoder allows for more accurate text generation by alleviating a crucial limitation in BERT-CTC,
wherein the model training is constrained on a word-level and domain-mismatched vocabulary used in BERT.

The key contributions of this work are summarized as
\begin{itemize}
    \item We introduce BERT-CTC, which efficiently incorporates linguistic knowledge from a pre-trained masked LM into the end-to-end ASR process.
    \item We propose BECTRA, an extension of BERT-CTC that effectively closes the gap between the text formats utilized in an end-to-end ASR model and pre-trained LM.
    \item We present probabilistic formulations of our proposed approaches and elucidate their close relationship to the conventional CTC and transducer-based models.
    \item We evaluate our models across various ASR tasks, which demonstrates the effectiveness irrespective of differences in the amount of training data, speaking styles, and languages. The codes and recipes have been made publicly available at \url{https://github.com/YosukeHiguchi/espnet/tree/bectra}.
    \item We perform thorough analyses to validate the efficacy of our models and propose several techniques to maximize their advantages for further improving ASR performance.
\end{itemize}

This paper builds upon our previous studies~\cite{higuchi2022bert,higuchi2023bectra}
by expanding on our findings in the following ways:
we present precise formulations of conventional end-to-end ASR approaches (Section~\ref{sec:background});
we provide a consistent description of the proposed BERT-CTC~\cite{higuchi2022bert} and BECTRA~\cite{higuchi2023bectra},
along with a comprehensive comparison of their formulations to the conventional approaches (Section~\ref{sec:proposed});
we provide detailed explanations of the relationship between our work and prior research (Section~\ref{sec:related_work});
we conduct experiments on various ASR tasks,
including additional experiments on low-resource and punctuation-preserved settings;
to enhance the effectiveness of our approach,
we explore the use of different pre-trained masked LMs and
the combination with shallow fusion (Sections~\ref{sec:experimental_setting} and~\ref{sec:results}); and
we further provide more in-depth analysis and investigation
to show the effectiveness of the proposed formulations and architectural designs (Section~\ref{sec:analyses}).

\section{Background: End-to-End ASR}
\label{sec:background}
To understand how the proposed approach utilizes a pre-trained masked LM (e.g., BERT~\cite{devlin2019bert}),
we start with a brief overview of the probabilistic formulations of conventional end-to-end ASR approaches,
focusing on CTC~\cite{graves2006connectionist,graves2014towards} and the transducer~\cite{graves2012sequence,graves2013speech}.

\subsection{Definition of End-to-End ASR}
\label{ssec:background_e2easr}
Let $O=(\bm{\mathrm{o}}_t\in\mathbb{R}^F| t=1,\cdots,T')$ be an input sequence of length $T'$, and
$W =( w_n\in\mathcal{V} | n=1,\cdots,N )$ be the corresponding output sequence of length $N$,
where $\bm{\mathrm{o}}_t$ is an $F$-dimensional acoustic feature at frame $t$, 
$w_n$ is an output token at position $n$, and $\mathcal{V}$ is a vocabulary.
In general,
the output length is much shorter than the input length (i.e., $N \ll T'$).
The goal of ASR is to identify the most probable output sequence $\hat{W}$
that matches the input sequence $O$:
\begin{equation}
    \hat{W} = \argmax_{W \in \mathcal{V}^{*}} p(W|O), \label{eq:asr}
\end{equation}
where $\mathcal{V}^{*}$ is a set of all possible token sequences.
End-to-end ASR aims to realize direct speech-to-text mapping ($O \mapsto W$)
by modeling the posterior distribution $p(W|O)$ using a single DNN.

\subsection{Conformer Encoder}
\label{ssec:background_conformer}
For the DNN architecture,
we adopt a Conformer-based model~\cite{gulati2020conformer}
consisting of a stack of $I$ identical encoder blocks.
The input audio sequence $O$ is embedded into a discriminative latent space as
\begin{equation}
    \label{eq:conformer}
    H = \text{ConformerEncoder}(O),
\end{equation}
where $H = (\bm{\mathrm{h}}_t \in \mathbb{R}^{d_{\mathsf{model}}} | t = 1,\cdots,T)$ is a sequence of $d_{\mathsf{model}}$-dimensional hidden vectors with length $T$ ($< T'$).
The $i$-th encoder block takes input as a previous sequence $H^{(i-1)} \in \mathbb{R}^{T \times d_{\mathsf{model}}}$ and
outputs $H^{(i)} \in \mathbb{R}^{T \times d_{\mathsf{model}}}$ as
\begin{align}
    \bar{H}^{(i)} &= H^{(i-1)} + \text{SelfAtten}(H^{(i-1)}), \label{eq:conformer_sa} \\
    H^{(i)} &= \bar{H}^{(i)\phantom{-1}} + \text{Conv}(\bar{H}^{(i)}), \label{eq:conformer_conv}
\end{align}
where $i \in \{1,\cdots,I\}$, and 
$\text{SelfAtten}(\cdot)$ and $\text{Conv}(\cdot)$ indicate the
multi-head self-attention and depthwise separable convolution modules, respectively.
$H^{(0)}$ is obtained by applying convolution down-sampling~\cite{hori2017advances} and positional encoding~\cite{vaswani2017attention} to $O$.
We denote $H$ in Eq.~\eqref{eq:conformer} as the final output of the blocks ($H = H^{(I)}$).
Note that in Eqs~\eqref{eq:conformer_sa} and~\eqref{eq:conformer_conv},
we omit layer normalization applied before each module and
the macaron-style feed-forward module for simplicity.

\begin{figure*}[t]
    \hspace{0.03cm}
    \begin{minipage}[b]{0.13\linewidth}
        \centering
        \includegraphics[height=4.0cm]{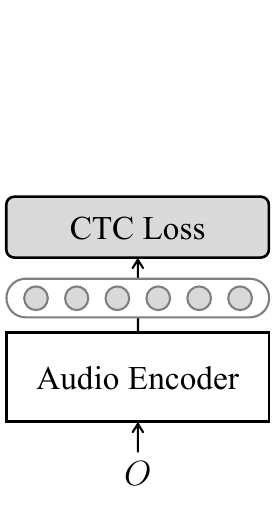}
        (a) CTC
    \end{minipage}
    \begin{minipage}[b]{0.24\linewidth}
        \centering
        \includegraphics[height=4.0cm]{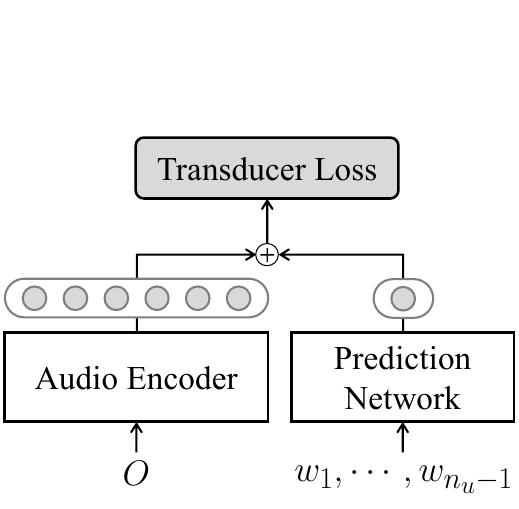}
        (b) Transducer
    \end{minipage}
    \begin{minipage}[b]{0.24\linewidth}
        \centering
        \includegraphics[height=4.0cm]{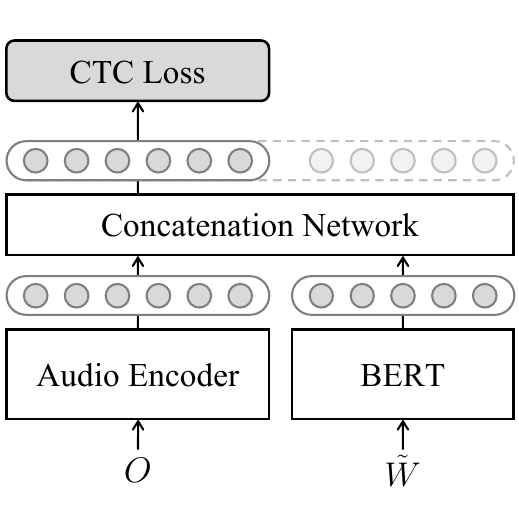}
        (c) BERT-CTC
    \end{minipage}
    \begin{minipage}[b]{0.36\linewidth}
        \centering
        \includegraphics[height=4.0cm]{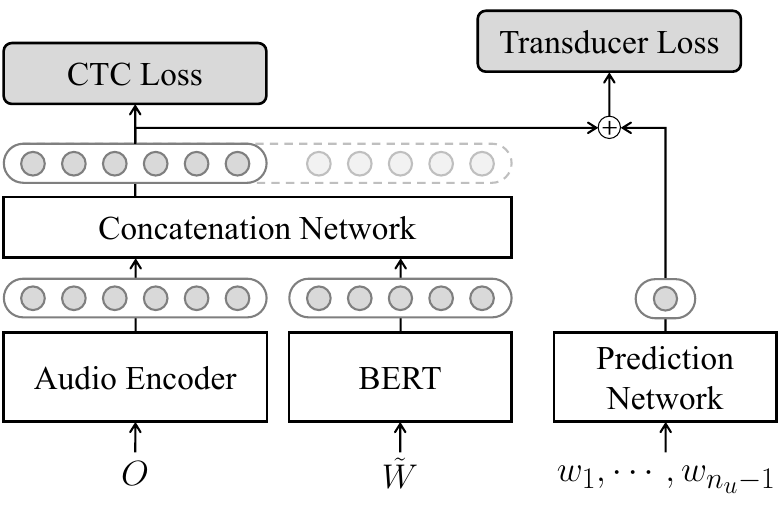}
        (d) BECTRA
    \end{minipage}
    \caption{Schematic comparisons between different model architectures for end-to-end ASR. All the models are described in relation to the CTC-based model (a). The transducer introduces a prediction network to capture causal dependencies between the outputs (b). BERT-CTC conditions CTC on contextualized linguistic representations that are obtained from BERT (c). BECTRA is an extension of BERT-CTC that incorporates a prediction network, leveraging the benefits of both the transducer framework and the usage of BERT (d).}
    \label{fig:e2easr}
\end{figure*}

\subsection{Connectionist Temporal Classification}
\label{ssec:background_ctc}
CTC~\cite{graves2006connectionist} formulates end-to-end ASR by
evaluating all possible alignments between an input sequence $O$ and the corresponding output sequence $W$.
To align the sequences at the frame level,
$W$ is augmented by permitting repeated occurrences of the same token and
inserting a blank symbol $\epsilon$ for representing ``no output token'' (e.g., silence).
Let $A=(a_t \in \mathcal{V} \cup \{\epsilon\} | t=1,\cdots,T)$ be an augmented sequence,
which we refer to as an \textit{alignment} between $O$ and $W$.

With the introduction of the frame-level alignment,
CTC factorizes the posterior distribution of $p(W|O)$ as
\begin{align}
    p_{\mathsf{ctc}}(W|O) &\approx \sum_{A \in \mathcal{B}_{\mathsf{ctc}}^{-1}(W)} p(W|A,\cancel{O}) p(A|O) \label{eq:p_ctc_W_O} \\
    &\approx \sum_{A \in \mathcal{B}_{\mathsf{ctc}}^{-1}(W)} p(A|O), \label{eq:p_ctc_W_O_approx}
\end{align}
where $\mathcal{B}_{\mathsf{ctc}}: A \mapsto W$ is the collapsing function that removes repeated tokens and blank symbols in $A$, and
$\mathcal{B}_{\mathsf{ctc}}^{-1}(W)$ is a set of all possible alignments that are compatible with $W$.
In Eq.~\eqref{eq:p_ctc_W_O}, CTC assumes a conditional independence of $O$,
which is indicated by the slash sign.
Furthermore, to obtain Eq.~\eqref{eq:p_ctc_W_O_approx},
$p(W|A)=1$ is assumed, as $W$ can be determined uniquely by the collapsing function.
The joint probability $p(A|O)$ in Eq.~\eqref{eq:p_ctc_W_O_approx} is further factorized by the probabilistic chain rule as
\begin{equation}
    p(A|O) \approx \prod_{t=1}^{T} p(a_t|\cancel{a_1,\cdots,a_{t-1}},O). \label{eq:p_ctc_A_O_approx}
\end{equation}
In Eq.~\eqref{eq:p_ctc_A_O_approx},
CTC makes a conditional independence assumption between output tokens, where
$p(A|O)$ is approximated as the product of token emission probabilities at each time frame.

The token emission probability $p(a_t|O)$ in Eq.~\eqref{eq:p_ctc_A_O_approx} is computed
using the embedded sequence $H$ from Eq.~\eqref{eq:conformer} as
\begin{equation}
    p(a_t|O) = \sigma(\bm{\mathrm{h}}_t) \in [0,1]^{|\mathcal{V}| + 1}, \label{eq:p_ctc_a_o}
\end{equation}
where $\sigma(\cdot)$ is a softmax layer.

\subsubsection{\textbf{Inference}}
\label{sssec:background_ctc_inference}
Substituting Eq.~\eqref{eq:p_ctc_W_O_approx} into Eq.~\eqref{eq:asr},
CTC estimates $\hat{W}$ using the best path decoding algorithm~\cite{graves2006connectionist}.
The most probable alignment $\hat{A}$ is first obtained
by concatenating the most active tokens at the each time frame in $H$: $\hat{a}_t=\argmax_{a_t} p(a_t|H)$.
$\hat{W}$ is then derived by applying the collapsing function to $\hat{A}$ as $\hat{W} = \mathcal{B}_{\mathsf{ctc}}(\hat{A})$.
Although not necessarily guaranteed to identify the most probable tokens,
this greedy algorithm has been empirically demonstrated to deliver satisfactory results,
particularly when the model is trained with the recent advanced modeling techniques~\cite{higuchi2021comparative}.

\subsubsection{\textbf{Training}}
\label{sssec:background_ctc_training}
The objective function of CTC is defined as the negative log-likelihood of Eq.~\eqref{eq:p_ctc_W_O_approx},
which is further expanded using Eq.~\eqref{eq:p_ctc_A_O_approx} as
\begin{align}
    \mathcal{L}_{\mathsf{ctc}} \triangleq -\log \sum_{A \in \mathcal{B}_{\mathsf{ctc}}^{-1}(W)} \prod_{t=1}^{T} p(a_t|O).
    \label{eq:L_ctc}
\end{align}
While Eq.~\eqref{eq:L_ctc} requires consideration of all possible $A$,
CTC efficiently computes it through dynamic programming (i.e., forward-backward algorithm).

\subsection{Transducer}
\label{ssec:background_transducer}
CTC estimates the distribution over alignments solely based on the speech input (i.e., Eq.~\eqref{eq:p_ctc_A_O_approx}),
leading to an inaccurate capture of the conditional dependence of output tokens, often referred to as the \textit{multimodality} problem~\cite{gu2018non}.
The transducer~\cite{graves2012sequence} addresses this problem by making each token prediction explicitly conditioned on the previous output tokens $(w_1,\cdots,w_{n-1})$.
Let $Z = (z_u\in\mathcal{V}\cup\{\epsilon\}|u=1,\cdots,T+N)$ be an alignment defined by the transducer,
which is slightly different from $A$ defined by CTC in that $Z$ takes into account the combined input and output states.

Similarly to Eq.~\eqref{eq:p_ctc_W_O_approx},
the transducer marginalizes the posterior distribution of $p(W|O)$ over all possible alignments as
\begin{equation}
    p_{\mathsf{tra}}(W|O) \approx \sum_{Z\in\mathcal{B}_{\mathsf{tra}}^{-1}(W)} p(Z|O), \label{eq:p_tra_W_O_approx}
\end{equation}
where $\mathcal{B}_{\mathsf{tra}}: Z \mapsto W$ is the collapsing function.
The joint probability $p(Z|O)$ in Eq.~\eqref{eq:p_tra_W_O_approx} is further factorized
\textit{without the conditional independence assumption} (cf. Eq.~\eqref{eq:p_ctc_A_O_approx}) as
\begin{align}
    p(Z|O) &= \prod_{u=1}^{T+N} p(z_{u}|z_1,\cdots,z_{u-1},O) \label{eq:p_tra_Z_O} \\
    &\approx \prod_{u=1}^{T+N} p(z_{u}|\underbrace{w_1,\cdots,w_{n_u-1}}_{= \mathcal{B}_{\mathsf{tra}}(z_1,\cdots,z_{u-1})},O), \label{eq:p_tra_Z_O_approx} 
\end{align}
where $n_u$ is the number of non-blank tokens predicted up to an output index of $u$.
From Eq.~\eqref{eq:p_tra_Z_O} to Eq.~\eqref{eq:p_tra_Z_O_approx},
the transducer approximates $(z_1,\cdots,z_{u-1})\approx(w_1,\cdots,w_{n_u-1})$,
which is reasonable because $W$ can be determined uniquely from $Z$ using the collapsing function.
The token emission probability $p(z_u|w_1,\cdots,w_{n_u-1},O)$ in Eq.~\eqref{eq:p_tra_Z_O_approx} is computed
using the embedded sequence $H$ from Eq.~\eqref{eq:conformer} as
\begin{align}
    p(z_{u}&|w_1,\cdots,w_{n_u-1},O) \nonumber \\
    &=\sigma(\text{JointNet}(\bm{\mathrm{h}}_t, \bm{\mathrm{q}}_{n_u})) \in [0,1]^{|\mathcal{V}|+1}, \label{eq:p_tra_z_W_O}\\
    \bm{\mathrm{q}}_{n_u} &= \text{PredictionNet} (w_1,\cdots,w_{n_u-1}) \in \mathbb{R}^{d_{\mathsf{model}}}. \label{eq:q_n_u}
\end{align}
In Eq.~\eqref{eq:p_tra_z_W_O}, $\text{JointNet}(\cdot)$ is a joint network that combines the audio and token representations,
$\bm{\mathrm{h}}_t$ and $\bm{\mathrm{q}}_{n_u}$, using a linear projection layer.
In Eq.~\eqref{eq:q_n_u}, $\text{PredictionNet}(\cdot)$ is a prediction network that
encodes the previous non-blank output tokens to a hidden vector $\bm{\mathrm{q}}_{n_u}$.
The introduction of the prediction network is the primary difference from CTC (Fig.~\ref{fig:e2easr}(a) vs.\ \ref{fig:e2easr}(b)),
which allows for explicit capture of causal dependencies in the outputs.

\subsubsection{\textbf{Inference}}
\label{sssec:background_tra_inference}
Based on Eqs.~\eqref{eq:asr} and~\eqref{eq:p_tra_W_O_approx},
the transducer identifies the most probable token sequence $\hat{W}$ using the beam search algorithm~\cite{graves2012sequence}.
The algorithm operates by searching through the possible hypotheses,
keeping the $B$-best hypotheses at each step based on their probability scores measured by Eq.~\eqref{eq:p_tra_z_W_O}.
We follow the implementation in~\cite{boyer2021study},
with a minor modification in skipping the initial prefix search part.

\subsubsection{\textbf{Training}}
\label{sssec:background_tra_training}
The transducer loss is defined by the negative log-likelihood of Eq.~\eqref{eq:p_tra_W_O_approx},
extended by Eq.~\eqref{eq:p_tra_Z_O_approx} as
\begin{equation}
    \mathcal{L}_{\mathsf{tra}} \triangleq -\log \sum_{Z \in \mathcal{B}_{\mathsf{tra}}^{-1}(W)}
    \prod_{u=1}^{T+N} p(z_{u}|w_{<n_u},O).
    \label{eq:L_tra}
\end{equation}
Similarly to the calculation of the CTC objective in Eq.~\eqref{eq:L_ctc},
the summation over alignments is efficiently implemented using dynamic programming.

\section{Integrating Pre-trained Masked Language Models into End-to-End ASR}
\label{sec:proposed}
We propose a novel approach to end-to-end ASR that utilizes a pre-trained masked LM in its formulation,
focusing on BERT~\cite{devlin2019bert} as a case in point.
Our proposed approach is designed to address the following key points:
\textit{How to make end-to-end ASR conditioned on BERT information}, and
\textit{how to bridge the gap between text processed in end-to-end ASR and BERT}.
The former is solved via \textbf{BERT-CTC},
which adapts BERT representations to explicitly condition CTC on linguistic contexts (Section~\ref{ssec:bertctc}).
The latter is tackled by \textbf{BE}RT-\textbf{C}TC-\textbf{T}ransducer (\textbf{BECTRA}) by
extending BERT-CTC to the transducer framework.
While the BERT-CTC formulation is restricted to the vocabulary used in BERT,
BECTRA overcomes this limitation by enabling the handling of different text formats and styles (Section~\ref{ssec:bectra}).

To provide a precise explanation of the proposed formulation,
we define output sequences that are tokenized using two varying vocabularies:
$W^{\mathsf{a}}=(w^{\mathsf{a}}_l \in \mathcal{V}^{\mathsf{a}}|l=1,\cdots,L)$ and
$W^{\mathsf{b}}=(w^{\mathsf{b}}_m \in \mathcal{V}^{\mathsf{b}}|m=1,\cdots,M)$, where
$\mathcal{V}^{\mathsf{a}}$ is a vocabulary constructed from ASR training text, and
$\mathcal{V}^{\mathsf{b}}$ is a vocabulary of BERT,
with the superscripts $\mathsf{a}$ and $\mathsf{b}$ indicating ASR and BERT, respectively.
Typically, $\mathcal{V}^{\mathsf{b}}$ consists of almost word-level tokens with a large subword vocabulary size,
while $\mathcal{V}^{\mathsf{a}}$ comprises smaller subword units (i.e., $|\mathcal{V}^{\mathsf{b}}| \gg |\mathcal{V}^{\mathsf{a}}|$).
Moreover, $\mathcal{V}^{\mathsf{b}}$ may contain written symbols,
including punctuation and casing,
whereas they are often disregarded in $\mathcal{V}^{\mathsf{a}}$.

\subsection{BERT-CTC}
\label{ssec:bertctc}
\noindent\textbf{Overview:}
In Fig.~\ref{fig:e2easr}, we compare BERT-CTC to the conventional CTC and transducer-based models (described in Sections~\ref{ssec:background_ctc} and~\ref{ssec:background_transducer}).
BERT-CTC leverages powerful contextualized representations from BERT
to make CTC's training and inference explicitly conditioned on linguistic information (Fig.~\ref{fig:e2easr}(a) vs.\ \ref{fig:e2easr}(c)).
BERT-CTC can be similar to the transducer in that
it fuses audio and token representations to estimate the distribution over alignments (Fig.~\ref{fig:e2easr}(b) vs.\ \ref{fig:e2easr}(c)).
However, by employing a concatenation network that attends to the full contexts of the input and output sequences,
BERT-CTC permits learning inner and inter-dependencies within and between the sequences,
facilitating the integration of information from the different modalities~\cite{fujita2020insertion}.

BERT-CTC formulates end-to-end ASR by introducing a partially masked ($\Leftrightarrow$ partially observed) sequence
$\tilde{W}^{\mathsf{b}}=(\tilde{w}^{\mathsf{b}}_m\in\mathcal{V}^{\mathsf{b}} | m=1,\cdots,M)$,
which is obtained by replacing some tokens in an output sequence $W^{\mathsf{b}}$ with a special mask token $\texttt{[MASK]}$.
Note that $\texttt{[MASK]}$ is included in the BERT vocabulary $\mathcal{V}^{\mathsf{b}}$.
An example pair of $W^{\mathsf{b}}$ and $\tilde{W}^{\mathsf{b}}$ can be
\begin{align}
    W^{\mathsf{b}} &= (\text{``Tokyo''}, \text{``is''}, \text{``the''}, \text{``capital''}, \text{``of''}, \text{``Japan''}, \text{``.''}), \nonumber \\
    \tilde{W}^{\mathsf{b}} &= (\text{``Tokyo''}, \text{\texttt{[MASK]}}, \text{``the''}, \text{\texttt{[MASK]}}, \text{``of''}, \text{``Japan''}, \text{``.''}). \nonumber
\end{align}
We obtain this masked sequence by applying masks to a ground-truth sequence during training or
a hypothesized sequence during inference.

Taking account of all possible masked sequences,
the posterior distribution of ASR, $p(W^{\mathsf{b}}|O)$, is factorized as
\begin{align}
    p_{\mathsf{bertctc}}(W^{\mathsf{b}}|O) &=\sum_{\tilde{W}^{\mathsf{b}} \in \mathcal{M}(W^{\mathsf{b}})} p(W^{\mathsf{b}},\tilde{W}^{\mathsf{b}}|O) \label{eq:p_bec_W_O_} \\
    &=\sum_{\tilde{W}^{\mathsf{b}} \in \mathcal{M}(W^{\mathsf{b}})}
    p(W^{\mathsf{b}}|\tilde{W}^{\mathsf{b}},O)
    p(\tilde{W}^{\mathsf{b}}|O), \label{eq:p_bec_W_O}
\end{align}
where $\mathcal{M}(W^{\mathsf{b}})$ covers $W^{\mathsf{b}}$ with all possible masking patterns.
In Eq.~\eqref{eq:p_bec_W_O}, we interpret $p(\tilde{W}^{\mathsf{b}}|O)$ as
a distribution of sequences that consist of unmasked (observed) tokens, which are readily recognizable from the speech input alone.
The other masked tokens, in contrast, are difficult to determine (e.g., homophones)
and require context from observed tokens, which is modeled by $p(W^{\mathsf{b}}|\tilde{W}^{\mathsf{b}},O)$.
We provide a more intuitive explanation of this interpretation in the inference section (Section~\ref{sssec:bertctc_inference}).

Similarly to Eq.~\eqref{eq:p_ctc_W_O},
$p(W^{\mathsf{b}}|\tilde{W}^{\mathsf{b}},O)$ in Eq.~\eqref{eq:p_bec_W_O}
is further factorized by introducing CTC alignments as
\begin{align}
    p(&W^{\mathsf{b}}|\tilde{W}^{\mathsf{b}},O) \nonumber \\
    &= \sum_{A^{\mathsf{b}} \in \mathcal{B}_{\mathsf{ctc}}^{-1}(W^{\mathsf{b}})} p(W^{\mathsf{b}},A^{\mathsf{b}}|\tilde{W}^{\mathsf{b}},O) \\
    &\approx \sum_{A^{\mathsf{b}} \in \mathcal{B}_{\mathsf{ctc}}^{-1}(W^{\mathsf{b}})} p(A^{\mathsf{b}}|W^{\mathsf{b}},\cancel{\tilde{W}^{\mathsf{b}}},O) p(W^{\mathsf{b}}|\tilde{W}^{\mathsf{b}},\cancel{O}), \label{eq:p_W_tW_O_approx}
\end{align}
where $A^{\mathsf{b}}=(a^{\mathsf{b}}_t\in\mathcal{V}^{\mathsf{b}}\cup\{\epsilon\}|t=1,\cdots,T)$ is an alignment corresponding to $W^{\mathsf{b}}$ with the BERT vocabulary.
In Eq.~\eqref{eq:p_W_tW_O_approx},
we make two conditional independence assumptions.
The first is that given $W^{\mathsf{b}}$ and $O$,
$\tilde{W}^{\mathsf{b}}$ is not required to determine $A^{\mathsf{b}}$.
This is reasonable because $W^{\mathsf{b}}$ already contains observed tokens in $\tilde{W}^{\mathsf{b}}$ and
is helpful in avoiding the combination of all possible masked sequences and alignments (i.e., the Cartesian product of $\mathcal{M}\times\mathcal{B}_{\mathsf{ctc}}^{-1}$).
The second is that given $\tilde{W}^{\mathsf{b}}$, $O$ is not required to determine $W^{\mathsf{b}}$.
We consider $p(W^{\mathsf{b}}|\tilde{W}^{\mathsf{b}})$ as a strong prior distribution
modeled by a pre-trained masked LM, e.g., BERT,
which can be achieved without the observation from $O$.
We empirically show that this assumption holds in Section~\ref{ssec:adapter_bert}.

The joint probability $p(A^{\mathsf{b}}|W^{\mathsf{b}},O)$ in Eq.~\eqref{eq:p_W_tW_O_approx} is factorized using the probabilistic chain rule as
\begin{equation}
    p(A^{\mathsf{b}}|W^{\mathsf{b}},O) \approx \prod_{t=1}^{T} p(a^{\mathsf{b}}_t|\cancel{a^{\mathsf{b}}_1,\cdots,a^{\mathsf{b}}_{t-1}},W^{\mathsf{b}},O). \label{eq:p_bec_A_W_O_approx}
\end{equation}
In Eq.~\eqref{eq:p_bec_A_W_O_approx},
we make the same conditional independence assumption as in CTC.
However, compared to Eq.~\eqref{eq:p_ctc_A_O_approx},
Eq.~\eqref{eq:p_bec_A_W_O_approx} is conditioned on an output sequence $W^{\mathsf{b}}$,
which enables explicit use of linguistic information to estimate the distribution over alignments.
This is somewhat similar to the transducer formulation in Eq.~\eqref{eq:p_tra_Z_O_approx},
but is different in that BERT-CTC attends to the whole context $(w^{\mathsf{b}}_1,\cdots,w^{\mathsf{b}}_M)$.

Substituting Eq.~\eqref{eq:p_bec_A_W_O_approx} into Eq.~\eqref{eq:p_W_tW_O_approx},
we model the product of $p(a^{\mathsf{b}}_t|W^{\mathsf{b}},O)$ and $p(W^{\mathsf{b}}|\tilde{W}^{\mathsf{b}})$ as
\begin{equation}
    \text{Eq.~\eqref{eq:p_W_tW_O_approx}} \triangleq \sum_{A^{\mathsf{b}} \in \mathcal{B}_{\mathsf{ctc}}^{-1}(W^{\mathsf{b}})} \prod_{t=1}^{T} p(a^{\mathsf{b}}_t|\text{BERT}(\tilde{W}^{\mathsf{b}}),O), \label{eq:p_W_tW_O_approx_bert}
\end{equation}
where $\text{BERT}(\cdot)$ indicates the final hidden states of BERT or any pre-trained masked LM,
representing the distribution of target sequences.\footnote{
As our formulation assumes $p(W^{\mathsf{b}}|\tilde{W}^{\mathsf{b}})$ to be a strong prior distribution of a masked LM,
we use BERT as a feature extractor for an output sequence without fine-tuning,
which has been reported to still be effective for several NLP tasks~\cite{peters2019tune,zhu2020incorporating,stappen2020cross}.
In Section~\ref{ssec:adapter_bert}, we provide empirical evidence that fine-tuning is not necessary for our proposed approach.
}
Eq.~\eqref{eq:p_W_tW_O_approx_bert} can be realized with a single differentiable model,
enabling the whole network to be trained end-to-end while being conditioned on BERT knowledge.
In Eq.~\eqref{eq:p_W_tW_O_approx_bert},
The token emission probability at each time frame is computed
using the audio sequence $H\in\mathbb{R}^{T \times d_{\mathsf{model}}}$ from Eq.~\eqref{eq:conformer} as
\begin{align}
    p(&a^{\mathsf{b}}_t|\text{BERT}(\tilde{W}^{\mathsf{b}}),O) \nonumber \\
    &= \sigma(\text{ConcatNet}_t((H,E))) \in [0, 1]^{|\mathcal{V}^{\mathsf{b}}|+1}, \label{eq:p_bec_a_tW_O} \\
    E &= \text{Linear}(\text{BERT}(\tilde{W}^{\mathsf{b}})) \in \mathbb{R}^{M \times d_{\mathsf{model}}}. \label{eq:E}
\end{align}
In Eq.~\eqref{eq:E}, $\text{Linear}(\cdot)$ maps the BERT outputs into
a sequence of $d_{\mathsf{model}}$-dimensional vectors $E$.
In Eq.~\eqref{eq:p_bec_a_tW_O},
$\text{ConcatNet}_t(\cdot)$ represents the $t$-th output of a concatenation network
that consists of a stack of Transformer self-attention layers~\cite{vaswani2017attention}.
Processing the concatenated sequence $(H,E)$ through this self-attention mechanism,
the model is capable of capturing dependencies within and between the audio and token sequences,
$H$ and $E$,
which we analyze in Section~\ref{ssec:attention_visualization}.

\subsubsection{\textbf{Inference}}
\label{sssec:bertctc_inference}
\begin{algorithm}[t]
    \caption{Decoding algorithm of BERT-CTC}
    \label{algo:bertctc_inference}
    \begin{algorithmic}[0]
        \algnotext{EndFor}
        \algnotext{EndFunction}
        \algrenewcommand\algorithmicindent{1.0em}
        \Function{DecodeBertctc}{$H$, $K$}
        \State \texttt{1}. Initialize a masked sequence $\tilde{W}'^{\mathsf{b}}$ with all mask tokens
        \For {$k=1$ to $K$}
            \State \texttt{2}. Forward $\text{BERT}(\tilde{W}'^{\mathsf{b}})$ and update $E$ with Eq.~\eqref{eq:E}
            \State \texttt{3}. Forward $\text{ConcatNet}(HE)$ and compute
            \Statex \qquad\qquad\ token emission probabilities with Eq.~\eqref{eq:p_bec_a_tW_O}
            \State \texttt{4}. Generate a hypothesis $\hat{W}^{\mathsf{b}}$ via best path decoding
            \State \texttt{5}. Compute $N_{\mathsf{mask}} = \lfloor |\hat{W}^{\mathsf{b}}| \cdot \frac{K - k}{K} \rfloor$
            \State \texttt{6}. Update $\tilde{W}'^{\mathsf{b}}$ by masking $N_{\mathsf{mask}}$ tokens in $\hat{W}^{\mathsf{b}}$
            \Statex \qquad\qquad\ with the lowest probability scores from Step \texttt{3}
        \EndFor
        \State \Return $\hat{W}^{\mathsf{b}}$, $E$
        \EndFunction
    \end{algorithmic}
\end{algorithm}
The most probable token sequence is estimated by solving Eq.~\eqref{eq:asr} for Eq.~\eqref{eq:p_bec_W_O} as
\begin{align}
    \hat{W}^{\mathsf{b}} &= \argmax_{W^{\mathsf{b}}} \sum_{\tilde{W}^{\mathsf{b}} \in \mathcal{M}(W^{\mathsf{b}})}
    p(W^{\mathsf{b}}|\tilde{W}^{\mathsf{b}},O)
    p(\tilde{W}^{\mathsf{b}}|O) \label{eq:bertctc_inference} \\
    &\approx \argmax_{W^{\mathsf{b}} } p(W^{\mathsf{b}} |\tilde{W}'^{\mathsf{b}}, O), \label{eq:bertctc_inference_approx} \\
    &\text{\hspace{0.46cm}where}\ \ \tilde{W}'^{\mathsf{b}}  = \argmax_{\tilde{W}^{\mathsf{b}}} p(\tilde{W}^{\mathsf{b}}|O). \label{eq:tilde_W_approx}
\end{align}
To obtain Eq.~\eqref{eq:bertctc_inference_approx},
we apply the Viterbi approximation to Eq.~\eqref{eq:bertctc_inference} in order to handle the intractable summation over all possible masked sequences.

The inference formulation with Eqs.~\eqref{eq:bertctc_inference_approx} and~\eqref{eq:tilde_W_approx}
can be viewed as the process of human speech recognition,
which involves ``top-down'' and ``bottom-up'' processing~\cite{mcclelland1986trace,norris1994shortlist}.
Determining $\tilde{W}'^{\mathsf{b}}$ in Eq.~\eqref{eq:tilde_W_approx} is analogous to bottom-up processing,
where the model analyzes the individual low-level sounds that make up words.
However,
certain words, particularly those with homophones, are difficult to identify solely from their sounds,
requiring higher-level linguistic information for accurate recognition.
This is solved via top-down processing in Eq.~\eqref{eq:bertctc_inference_approx},
where the conditioning from $\tilde{W}'^{\mathsf{b}}$
enables the model to leverage linguistic knowledge, context, and anticipations
for identifying words from low-level sounds.

To solve Eqs.~\eqref{eq:bertctc_inference_approx} and~\eqref{eq:tilde_W_approx},
we design a fill-mask-style decoding algorithm based on mask-predict~\cite{ghazvininejad2019mask} and CTC inference,
which is based on~\cite{chan2020imputer,higuchi2020mask,higuchi2021improved}.
Algorithm~\ref{algo:bertctc_inference}, consisting of Steps \texttt{1} to \texttt{6}, describes the proposed algorithm.
At the beginning of decoding,
a masked sequence $\tilde{W}'^{\mathsf{b}}$ is initialized by
replacing all token positions with the mask token $\texttt{[MASK]}$ (Step \texttt{1}).\footnote{This requires predicting the target length beforehand~\cite{gu2018non}, which we obtain from intermediate predictions from the encoder (see Section~\ref{ssec:decoding_configuration} for details).}
The algorithm then proceeds to generate a hypothesis by
gradually filling in the masked tokens over $K$ iterations.
At each iteration $k \in \{1,\cdots,K\}$,
the current masked sequence $\tilde{W}'^{\mathsf{b}}$ is fed into BERT to obtain contextual embeddings $E$,
as defined by Eq.~\eqref{eq:E} (Step \texttt{2}).
The encoder output $H$ from Eq.~\eqref{eq:conformer} and $E$ are concatenated and input into the concatenation network,
which computes the framewise probability $p(a_t^{\mathsf{b}}|\text{BERT}(\tilde{W}'^{\mathsf{b}}),O)$
as in Eq.~\eqref{eq:p_bec_a_tW_O} (Step \texttt{3}).
Using the probabilities computed at each frame,
a hypothesis $\hat{W}^{\mathsf{b}}$ is generated through best path decoding,
in the same manner as in Section~\ref{sssec:background_ctc_inference} (Step \texttt{4}).
The number of tokens that will be masked $N_{\mathsf{mask}}$ is determined by a linear decay function as
$N_{\mathsf{mask}} = \lfloor |\hat{W}^{\mathsf{b}}| \cdot \frac{K - k}{K} \rfloor$ (Step \texttt{5}), e.g.,
if $K$ is set to five, $N_{\mathsf{mask}}$ decreases by 20\% at each iteration.
The masked sequence $\tilde{W}'^{\mathsf{b}}$ is updated by
replacing $N_{\mathsf{mask}}$ tokens in the hypothesis $\hat{W}^{\mathsf{b}}$ with the mask token $\texttt{[MASK]}$ (Step \texttt{6}).
Here, tokens are selected for masking according to their confidence scores,
which are measured by calculating the output probability of each token.
Using the framewise probabilities from Step \texttt{3},
the output probability for a token $\hat{w}^{\mathsf{b}}_n \in \hat{W}^{\mathsf{b}}$ is derived as
\begin{align}
    p(&w_n^{\mathsf{b}} = \hat{w}_n^{\mathsf{b}}|\text{BERT}(\tilde{W}'^{\mathsf{b}}), O) \nonumber \\
    &= \text{max}\left(\left\{p(a_t^{\mathsf{b}} = \hat{w}_n^{\mathsf{b}}|\text{BERT}(\tilde{W}'^{\mathsf{b}}), O) | t \in \mathcal{T}_n \right\}\right),
    \label{eq:token_prob}
\end{align}
where $\mathcal{T}_n$ is a set of frame indices that correspond to the $n$-th token $\hat{w}^{\mathsf{b}}_n$
after applying the collapsing function.
With Eq.~\eqref{eq:token_prob}, $N_{\mathsf{mask}}$ tokens with the lowest  probability scores are masked.

In the first iteration (i.e., $k = 1$),
the model generates a hypothesis solely based on the speech input,
without any linguistic cues from the output tokens, which are all masked.
This can be aligned with the concept of bottom-up processing as formulated by Eq.~\eqref{eq:tilde_W_approx}.
As the iterations proceed (i.e., $1 < k \le K$), the output tokens become gradually observable,
providing additional linguistic information for generating a more precise hypothesis.
This can be interpreted as top-down processing as formulated by Eq.~\eqref{eq:bertctc_inference_approx}.

\subsubsection{\textbf{Training}}\label{sssec:bertctc_training}
The BERT-CTC objective is defined by the negative log-likelihood of Eq.~\eqref{eq:p_bec_W_O}:
\begin{equation}
    -\log \sum_{\tilde{W}^{\mathsf{b}} \in \mathcal{M}(W^{\mathsf{b}})}
    p(W^{\mathsf{b}}|\tilde{W}^{\mathsf{b}},O) p(\tilde{W}^{\mathsf{b}}|O),
\end{equation}
which is further expanded using Eqs.~\eqref{eq:p_W_tW_O_approx} and~\eqref{eq:p_bec_A_W_O_approx} as
\begin{equation}
    \hspace{-0.5cm}= -\log \sum_{\substack{\tilde{W}^{\mathsf{b}} \in \mathcal{M}(W^{\mathsf{b}}) \\ A^{\mathsf{b}} \in \mathcal{B}_{\mathsf{ctc}}^{-1}(W^{\mathsf{b}})}}
    p(A^{\mathsf{b}}|W^{\mathsf{b}},O) p(W^{\mathsf{b}}|\tilde{W}^{\mathsf{b}}) p(\tilde{W}^{\mathsf{b}}|O) \!
\end{equation}
\begin{equation}
    \approx - \log \mathbb{E}_{\tilde{W}^{\mathsf{b}} \sim \mathcal{M}'(W^{\mathsf{b}})}
    \!\Bigg[\sum_{A^{\mathsf{b}} \in \mathcal{B}_{\mathsf{ctc}}^{-1}(W^{\mathsf{b}})}
    \!p(A^{\mathsf{b}}|W^{\mathsf{b}},O) p(W^{\mathsf{b}}|\tilde{W}^{\mathsf{b}}) \Bigg].
    \label{eq:L_bec_exp}
\end{equation}
To obtain Eq.~\eqref{eq:L_bec_exp},
we approximate the intractable marginalization over $\tilde{W}^{\mathsf{b}}$ as
expectation with respect to the sampling distribution $\mathcal{M}'(W^{\mathsf{b}})$,
which is calculated on the probability distribution of $p(\tilde{W}^{\mathsf{b}}|O)$.
The upper bound of Eq.~\eqref{eq:L_bec_exp} can be derived by applying Jensen's inequality as
\begin{equation}
    \approx - \mathbb{E}_{\tilde{W}^{\mathsf{b}} \sim \mathcal{M}'(W^{\mathsf{b}})}
    \Bigg[ \log\!\sum_{A^{\mathsf{b}} \in \mathcal{B}_{\mathsf{ctc}}^{-1}(W^{\mathsf{b}})}
    \!p(A^{\mathsf{b}}|W^{\mathsf{b}},O) p(W^{\mathsf{b}}|\tilde{W}^{\mathsf{b}}) \Bigg].
    \label{eq:L_bec_upper}
\end{equation}
Substituting Eqs.~\eqref{eq:p_bec_A_W_O_approx} and~\eqref{eq:p_W_tW_O_approx_bert} into Eq.~\eqref{eq:L_bec_upper},
the loss for BERT-CTC training is defined as
\begin{align}
    &\mathcal{L}_{\mathsf{bertctc}} \triangleq  \nonumber \\
    &- \mathbb{E}_{\tilde{W}^{\mathsf{b}} \sim \mathcal{M}'(W^{\mathsf{b}})}
    \Bigg[\log \sum_{A^{\mathsf{b}} \in \mathcal{B}_{\mathsf{ctc}}^{-1}(W^{\mathsf{b}})}
    \prod_{t=1}^{T} p(a^{\mathsf{b}}_t|\text{BERT}(\tilde{W}^{\mathsf{b}}),O)\Bigg].
    \label{eq:L_bec}
\end{align}
In comparison to the CTC objective described in Eq.~\eqref{eq:L_ctc},
each token prediction in Eq.~\eqref{eq:L_bec} is explicitly conditioned on the contextualized embeddings from BERT.
This allows an explicit consideration of the contextual dependencies among token predictions
while retaining the efficient optimization strategy as in CTC.

For the sampling process of $\tilde{W}^{\mathsf{b}}$ in Eq.~\eqref{eq:L_bec},
we use random sampling from a uniform distribution to approximate the probability distribution of $\mathcal{M}'(W^{\mathsf{b}})$,
for the sake of simplicity.
We adopt a strategy similar to the one employed in~\cite{ghazvininejad2019mask}.
We first sample a random number $N_{\mathsf{mask}}$ from a uniform distribution
ranging between one and the target sequence length $M$,
i.e., $N_{\mathsf{mask}} \sim \text{Uniform}(1, M)$.
Then, we randomly select $N_{\mathsf{mask}}$ tokens from a ground-truth sequence and replace them with \texttt{[MASK]}.

\subsection{BERT-CTC-Transducer (BECTRA)}
\label{ssec:bectra}
BERT-CTC progressively conditions CTC on contextual linguistic information by
gradually predicting tokens and updating BERT embeddings correspondingly.
This BERT-based refinement requires the model to work with the BERT's text format,
which has the vocabulary $\mathcal{V}^{\mathsf{b}}$ that can be too large for ASR training, and
could lead to a mismatch against the target ASR domain.
BERT-CTC-Transducer (BECTRA) is designed to extend BERT-CTC for handling such mismatches
while still utilizing BERT knowledge to enhance ASR performance.

\noindent\textbf{Overview:}
Figure~\ref{fig:e2easr} presents a comparison between BECTRA,
the conventional transducer-based model (described in Section~\ref{ssec:background_transducer}), and
BERT-CTC (introduced in Section~\ref{ssec:bertctc}).
BECTRA formulates end-to-end ASR based on BERT-CTC,
using the output of the concatenation network for calculating the transducer loss (Fig.~\ref{fig:e2easr}(c) vs.\ \ref{fig:e2easr}(d)).
Here, the joint and prediction networks are trained with an ASR-specific vocabulary $\mathcal{V}^{\mathsf{a}}$,
which allows for more suitable end-to-end ASR training without being limited by the BERT vocabulary $\mathcal{V}^{\mathsf{b}}$.
Hence, unlike $p_{\mathsf{bertctc}}(W^{\mathsf{b}}|O)$ in Eq.~\eqref{eq:p_bec_W_O_},
BECTRA utilizes $W^{\mathsf{a}}$ tokenized by $\mathcal{V}^{\mathsf{a}}$ as its target sequence.

Similarly to Eq.~\eqref{eq:p_bec_W_O},
BECTRA formulates E2E-ASR by marginalizing the posterior distribution of $p(W^{\mathsf{a}}|O)$ over all possible masked sequences as
\begin{equation}
    p_{\mathsf{bectra}}(W^{\mathsf{a}}|O) = \sum_{\tilde{W}^{\mathsf{b}}\in\mathcal{M}(W^{\mathsf{b}})} p(W^{\mathsf{a}}|\tilde{W}^{\mathsf{b}},O) p(\tilde{W}^{\mathsf{b}}|O).
    \label{eq:p_bectra_W_O}
\end{equation}
In Eq.~\eqref{eq:p_bectra_W_O},
$\tilde{W}^{\mathsf{b}}$ is obtained by masking a sequence in the BERT unit $W^{\mathsf{b}}$ ($\neq W^{\mathsf{a}}$).
Similarly to Eq.~\eqref{eq:p_tra_W_O_approx},
$p(W^{\mathsf{a}}|\tilde{W}^{\mathsf{b}},O)$ in Eq.~\eqref{eq:p_bectra_W_O} is factorized
by considering all possible alignments of the transducer as
\begin{align}
    p(&W^{\mathsf{a}}|\tilde{W}^{\mathsf{b}},O) \nonumber \\
    &= \sum_{Z^{\mathsf{a}}\in\mathcal{B}_{\mathsf{tra}}^{-1}(W^{\mathsf{a}})}
    p(W^{\mathsf{a}}, Z^{\mathsf{a}}|\tilde{W}^{\mathsf{b}},O) \\
    &\approx \sum_{Z^{\mathsf{a}}\in\mathcal{B}_{\mathsf{tra}}^{-1}(W^{\mathsf{a}})}
    p(Z^{\mathsf{a}}|W^{\mathsf{a}},\cancel{\tilde{W}^{\mathsf{b}}},O) p(W^{\mathsf{a}}|\tilde{W}^{\mathsf{b}}, \cancel{O}),
    \label{eq:p_bectra_W_tW_O}
\end{align}
where $Z^{\mathsf{a}}=(z^{\mathsf{a}}_u \in \mathcal{V}^{\mathsf{a}}\cup\{\epsilon\}|u = 1,\cdots,T + L)$
is an alignment corresponding to $W^{\mathsf{a}}$ with the ASR vocabulary,
as defined by the transducer (see Eq.~\eqref{eq:p_tra_W_O_approx}).
In Eq.~\eqref{eq:p_bectra_W_tW_O},
we use the same approximations employed in Eq.~\eqref{eq:p_W_tW_O_approx}.
The joint probability $p(Z^{\mathsf{a}}|W^{\mathsf{a}},O)$ in Eq.~\eqref{eq:p_bectra_W_tW_O} is further factorized by the probabilistic chain rule
\textit{without a conditional independence assumption} (cf. Eq.~\eqref{eq:p_bec_A_W_O_approx}) as
\begin{align}
    p(&Z^{\mathsf{a}}|W^{\mathsf{a}},O) \nonumber \\
    &= \sum_{Z^{\mathsf{a}}\in\mathcal{B}_{\mathsf{tra}}^{-1}(W^{\mathsf{a}})} \prod_{u=1}^{T+L} p(z^{\mathsf{a}}_u|z^{\mathsf{a}}_1,\cdots,z^{\mathsf{a}}_{u-1},W^{\mathsf{a}},O), \\
    &\approx \sum_{Z^{\mathsf{a}}\in\mathcal{B}_{\mathsf{tra}}^{-1}(W^{\mathsf{a}})} \prod_{u=1}^{T+L} p(z^{\mathsf{a}}_u|\underbrace{w^{\mathsf{a}}_1,\cdots,w^{\mathsf{a}}_{l_{u}-1}}_{=\mathcal{B}_{\mathsf{tra}}(z^{\mathsf{a}}_1,\cdots,z^{\mathsf{a}}_{u-1})},W^{\mathsf{a}},O),
    \label{eq:p_bectra_W_tW_O_2b}
\end{align}
where $l_u$ is the number of non-blank tokens predicted up to an index of $u$.
Eq.~\eqref{eq:p_bectra_W_tW_O_2b} assumes $(z^{\mathsf{a}}_1,\cdots,z^{\mathsf{a}}_{u-1}) \approx (w^{\mathsf{a}}_1,\cdots,w^{\mathsf{a}}_{l_u-1})$,
using the same approximation as the transducer in Eq.~\eqref{eq:p_tra_Z_O_approx}.
Similarly to the BERT-CTC formulation in Eq.~\eqref{eq:p_W_tW_O_approx_bert},
BECTRA models Eq.~\eqref{eq:p_bectra_W_tW_O} using Eq.~\eqref{eq:p_bectra_W_tW_O_2b} as
\begin{equation}
    \text{Eq.~\eqref{eq:p_bectra_W_tW_O}} \triangleq \sum_{Z^{\mathsf{a}}\in\mathcal{B}_{\mathsf{tra}}^{-1}(W^{\mathsf{a}})}\prod_{u=1}^{T+L}
    p(z^{\mathsf{a}}_u|w^{\mathsf{a}}_{<l_{u}},\text{BERT}(\tilde{W}^{\mathsf{b}}),O), \label{eq:p_bectra_W_tW_O_3}
\end{equation}
where we employ BERT to model $p(W^{\mathsf{a}}|\tilde{W}^{\mathsf{b}})$.
This is a reasonable approximation because both $W^{\mathsf{b}}$ and $W^{\mathsf{a}}$ represent the same target sentence.
Thus, $W^{\mathsf{a}}$ can be derived easily by first converting $W^{\mathsf{b}}$ into a word sequence and
then tokenizing it using $\mathcal{V}^{\mathsf{a}}$.
The token emission probability in Eq.~\eqref{eq:p_bectra_W_tW_O_3} is computed using
the audio sequence $H$ from Eq.~\eqref{eq:conformer} and BERT output $E$ similar to Eq.~\eqref{eq:p_bec_a_tW_O} as
\begin{align}
    &\hspace{-0.12cm}p(z^{\mathsf{a}}_u|w^{\mathsf{a}}_{<l_{u}},\text{BERT}(\tilde{W}^{\mathsf{b}}),O) \nonumber \\ &\hspace{-0.12cm}=\sigma(\text{JointNet}(\text{ConcatNet}_t((H,E)),\bm{\mathrm{q}}_{l_u}^{\mathsf{a}})) \in [0,1]^{|\mathcal{V}^{\mathsf{a}}|+1},\! \label{eq:p_bectra_at} \\
    &\hspace{-0.12cm}\bm{\mathrm{q}}_{l_u}^{\mathsf{a}} = \text{PredictionNet} (w^{\mathsf{a}}_1,\cdots,w^{\mathsf{a}}_{l_u-1}) \in \mathbb{R}^{d_{\mathsf{model}}}. \label{eq:q_l_u}
\end{align}
The network architecture is almost identical to the transducer presented in Eqs.~\eqref{eq:p_tra_z_W_O} and~\eqref{eq:q_n_u}
but it differs in that the joint network takes the output of the concatenation network as its input.
This enables the integration of BERT knowledge into the transducer-based model.
By adopting the prediction network,
BECTRA can explicitly capture the causal dependencies between output tokens,
leading to better sequence modeling.
This is another key advantage of BECTRA compared to BERT-CTC,
beyond the use of ASR-specific vocabulary,
which we discuss in Section~\ref{ssec:bectra_bert}.

\subsubsection{\textbf{Inference}}
\begin{algorithm}[t]
    \caption{Decoding algorithm of BECTRA}
    \label{algo:bectra_inference}
    \begin{algorithmic}[0]
        \algnotext{EndFunction}
        \Function{DecodeBectra}{$H$, $K$, $B$}
        \State \texttt{1}. Perform \Call{DecodeBertctc}{$H$, $K$} and obtain $E$
        \State \texttt{2}. Generate a hypothesis $\hat{W}^{\mathsf{a}}$ via beam search decoding
        \Statex \qquad\qquad\ with a beam size of $B$, using output probabilities
        \Statex \qquad\qquad\ computed with Eqs.~\eqref{eq:p_bectra_at} and~\eqref{eq:q_l_u}
        \State \Return $\hat{W}^{\mathsf{a}}$
        \EndFunction
    \end{algorithmic}
\end{algorithm}
Algorithm~\ref{algo:bectra_inference} shows the inference algorithm of BECTRA, which includes Steps \texttt{1} and \texttt{2}.
The algorithm is implemented with BERT-CTC decoding (see Section~\ref{sssec:bertctc_inference})
followed by beam-search decoding of the transducer (see Section~\ref{sssec:background_tra_inference}).
BERT-CTC decoding provides the model with a fully contextualized BERT output $E$,
which is obtained from the final hypothesis estimated by the iterative refinement (Step \texttt{1}).
To find the optimal sequence with the highest sequence-level generation probability,
beam-search decoding is performed using the token emission probabilities computed from Eq.~\eqref{eq:p_bectra_at} (Step \texttt{2}).
With this combined inference algorithm,
BECTRA can leverage the BERT's ability to capture bi-directional context in an output sequence,
providing the benefit of non-autoregressive decoding.
Furthermore, transducer-based decoding enables the model to refine a sequence in an autoregressive manner,
utilizing a more appropriate output unit for performing ASR.

\subsubsection{\textbf{Training}}
The transducer loss of BECTRA is defined by substituting Eq.~\eqref{eq:p_bectra_W_tW_O_3} into Eq.~\eqref{eq:p_bectra_W_O}
and following the same derivation process as Eq.~\eqref{eq:L_bec}, resulting in
{\small
\begin{align}
    &\mathcal{L}'_{\mathsf{bectra}} \triangleq \nonumber \\
    &-\mathbb{E}_{\tilde{W}^{\mathsf{b}} \sim \mathcal{M}(W^{\mathsf{b}})} \Bigg[ \log \sum_{Z^{\mathsf{a}}\in\mathcal{B}_{\mathsf{tra}}^{-1}(W^{\mathsf{a}})}\!\prod_{u=1}^{T+L} \!p(z^{\mathsf{a}}_u|w^{\mathsf{a}}_{<l_{u}},\text{BERT}(\tilde{W}^{\mathsf{b}}),O) \Bigg], \label{eq:L_bectra_}
\end{align}
}%
where the summation over $Z^{\mathsf{a}}$ can be efficiently computed using the same approach as in transducer training (see Section~\ref{sssec:background_tra_training}).
The sampling strategy for $\tilde{W}^{\mathsf{b}}$ is described in BERT-CTC training (see Section~\ref{sssec:bertctc_training}).
The objective function of BECTRA is defined by combining $\mathcal{L}_{\mathsf{bertctc}}$ from Eq.~\eqref{eq:L_bec} and $\mathcal{L}'_{\mathsf{bectra}}$ from Eq.~\eqref{eq:L_bectra_} as
\begin{equation}
    \mathcal{L}_{\mathsf{bectra}} = (1 - \lambda) \mathcal{L}_{\mathsf{bertctc}} + \lambda \mathcal{L}'_{\mathsf{bectra}},
    \label{eq:L_bectra}
\end{equation}
where $\lambda \in (0,1)$ is a tunable parameter.

\subsection{Overall Comparison of End-to-End ASR Formulations}
\label{ssec:formulations}
\begin{table}[t]
    \centering
    \caption{Comparisons between end-to-end ASR formulations for modeling distribution over alignments}
    \label{tb:formulations}
    \begin{tabular}{l@{\hspace{2.5\tabcolsep}}l@{\hspace{2.5\tabcolsep}}}
        \toprule
        \textbf{Model} & \textbf{Formulation} \\
        \midrule
        \multirow{1}{*}[-2pt]{CTC} & $\displaystyle \sum_{A \in \mathcal{B}_{\mathsf{ctc}}^{-1}(W)} \prod_{t=1}^{T} p(a_t|O)$ \\
        \cmidrule{1-2}
        \multirow{1}{*}[-2pt]{Transducer} & $\displaystyle \sum_{Z \in \mathcal{B}_{\mathsf{tra}}^{-1}(W)} \prod_{u=1}^{T+N} p(z_{u}|w_{<n_u},O)$ \\
        \cmidrule{1-2}
        \multirow{1}{*}[-2pt]{BERT-CTC} & $\displaystyle \sum_{A \in \mathcal{B}_{\mathsf{ctc}}^{-1}(W)} \prod_{t=1}^{T} p(a_t|\text{BERT}(\tilde{W}), O)$ \\
        \cmidrule{1-2}
        \multirow{1}{*}[-2pt]{BECTRA} & $\displaystyle \sum_{Z \in \mathcal{B}_{\mathsf{tra}}^{-1}(W)} \prod_{u=1}^{T+N} p(z_{u}|w_{<n_u}, \text{BERT}(\tilde{W}), O)$ \\
        \bottomrule
    \end{tabular}
\end{table}
As summarized in Table~\ref{tb:formulations},
the key difference between the end-to-end ASR formulations discussed thus far 
lies in how the distribution over alignments is calculated.
CTC estimates the distribution based solely on the speech input $O$,
assuming that the output tokens are independent of one another.
The transducer conditions each token prediction explicitly on the preceding non-blank tokens $w_{<n_u}$,
introducing the prediction and joint networks.
BERT-CTC achieves similar conditioning using BERT's contextualized embeddings $\text{BERT}(\tilde{W})$
through the concatenation network.
BECTRA is conditioned on both $w_{<n_u}$ and $\text{BERT}(\tilde{W})$,
enabling a model to benefit from the information provided by both sources.

\section{Additional Related Work}
\label{sec:related_work}
In this section, we further clarify the position of our research
in relation to other relevant topics in end-to-end ASR,
focusing on masked LM-based modeling and external LM integration.

\subsection{End-to-End ASR and Masked Language Modeling}
Conditional masked LM (CMLM)~\cite{ghazvininejad2019mask}, one of the successful approaches in non-autoregressive neural machine translation,
has been introduced to solve end-to-end ASR.
CMLM utilizes an encoder-decoder structure,
wherein its decoder is trained with the masked LM objective~\cite{devlin2019bert}
while being conditioned on the encoder outputs through the cross-attention mechanism.
Audio-CMLM~\cite{chen2021non} employs CMLM to enable non-autoregressive end-to-end ASR by
conditioning the decoder on audio information to learn the fill-mask process.
By combining CMLM-based modeling with CTC,
Imputer~\cite{chan2020imputer} and Mask-CTC~\cite{higuchi2020mask,higuchi2021improved} extend the mask-predict algorithm
to refine either a frame-level or token-level sequence predicted in the CTC framework.
Several studies have trained CMLM as an error-correction model for predictions generated by an end-to-end ASR system~\cite{futami2022non,fan2022acoustic}.

Our approach of incorporating masked language modeling with the CTC and transducer frameworks is relevant to the above studies.
However, it differs in that we aim to leverage the pre-existing knowledge acquired by a pre-trained masked LM to enhance end-to-end ASR performance.

\subsection{Language Model Integration for End-to-End ASR}
There is a line of previous studies that have explored the integration of an LM,
e.g., recurrent neural network (RNN)-LM, into end-to-end ASR systems.
In this context, the LM is trained on external in-domain text data that pertains to a specific ASR task.
Shallow fusion has been the predominant method~\cite{hannun2014deep,gulcehre2015using,chorowski2017towards,kannan2018analysis},
which linearly interpolates the output probabilities from an end-to-end ASR model and external LM.
Deep fusion~\cite{gulcehre2015using} is a more structured approach,
where an end-to-end ASR model is jointly trained with an external LM
to learn the optimal combination of audio and linguistic information.
Cold fusion~\cite{sriram2018cold} and component fusion~\cite{shan2019component} have improved deep fusion by
incorporating a gating mechanism that enables a more sophisticated combination of the two models.

Our approach shares similarities with cold fusion by combining
an end-to-end ASR model and pre-trained masked LM 
using the self-attention mechanism to selectively merge audio and linguistic representations.
However, we focus on exploring
how the versatile linguistic knowledge acquired from large-scale pre-trained LMs (i.e., BERT)
can be utilized to improve end-to-end ASR.
Additionally,
we demonstrate that the conventional LM fusion technique is applicable to our approach,
allowing for the incorporation of a domain-specific RNN-LM to further enhance performance.

\section{Experimental Setting}
\label{sec:experimental_setting}
We used the ESPnet toolkit~\cite{watanabe2018espnet,watanabe20212020} for conducting the experiments.
All the codes and recipes used in our experiments are
made publicly available.\footnote{\url{https://github.com/YosukeHiguchi/espnet/tree/bectra}}

\subsection{Data}
\label{ssec:experimental_setting_data}
\begin{table}[t]
    \centering
    \caption{Dataset descriptions}
    \label{tb:datasets}
    \begin{tabular}{l@{\hspace{1.3\tabcolsep}}c@{\hspace{1.5\tabcolsep}}c@{\hspace{1.5\tabcolsep}}c@{\hspace{1.5\tabcolsep}}c@{\hspace{1.5\tabcolsep}}}
        \toprule
        \textbf{Dataset} & \textbf{Hours} & \textbf{Language} & \textbf{Speech Style} & \textbf{Text Style} \\
        \midrule
        \multirow{2}{*}[0pt]{LibriSpeech~\cite{panayotov2015librispeech}} & 100 & English & Read & Normalized \\
        & 960 & English & Read & Normalized \\
        \cmidrule{1-5}
        Libri-Light~\cite{kahn2020libri} & \phantom{0}10 & English & Read & Normalized \\
        \cmidrule{1-5}
        TED-LIUM2~\cite{rousseau2014enhancing} & 210 & English & Spontaneous & Normalized \\
        \cmidrule{1-5}
        AISHELL-1~\cite{bu2017aishell} & 170 & Mandarin & Read & Normalized \\
        \cmidrule{1-5}
        CoVoST2~\cite{wang2021covost} & 430 & English & Read & Punct./Casing \\
        \bottomrule
    \end{tabular}
\end{table}
We used the corpora listed in Table~\ref{tb:datasets},
which comprised different quantities of data, languages, and speech and text styles.
LibriSpeech~\cite{panayotov2015librispeech} consists of utterances derived from read English audiobooks.
In addition to the full 960-hour training set (LS-960),
we trained models on the \textit{train-clean-100} subset (LS-100)
for conducting additional investigations and analyses.
We also used the 10-hour training set provided by Libri-Light~\cite{kahn2020libri} (LL-10),
which is a low-resource subset extracted from LS-960.
TED-LIUM2 (TED2) ~\cite{rousseau2014enhancing} contains utterances from English TED Talks.
AISHELL-1 (AS1)~\cite{bu2017aishell} is a multi-domain Mandarin corpus that covers a range of common applications,
such as prompts for smart speakers.
CoVoST2 (CV2)~\cite{wang2021covost} is a corpus for speech translation tasks,
which is based on the Common Voice project~\cite{ardila2020common}.
We used CV2 for an English ASR task by exclusively using source speech-text data from the ``En$\rightarrow$X'' task.
For each corpus,
we used the standard development and test sets for tuning hyper-parameters and evaluating performance, respectively.

Notice that all corpora except CV2 only provide normalized transcriptions
in which punctuation is removed and casing is standardized to upper or lower case.
This potentially limits the capabilities of BERT,
which is often trained on written-form text with punctuation and casing preserved.
In contrast, CV2 provides unnormalized transcriptions with a decent amount of ASR training data,
which makes it an ideal resource for evaluating the effectiveness of the proposed approach.

We used SentencePiece~\cite{kudo2018subword} to construct subword vocabularies from ASR transcriptions
in order to obtain the ASR vocabulary $\mathcal{V}^{\mathsf{a}}$.
The vocabulary sizes were set to $300$, $5\text{k}$, $100$, $500$, and $500$
for LS-100, LS-960, LL-10, TED2, and CV2, respectively.
For AS1, we used character-level tokenization with $4231$ Chinese characters.
It should be noted that before extracting subwords or characters, the ASR transcriptions were normalized
regardless of the corpus.

\subsection{Model and Network Architecture}
We evaluated end-to-end ASR models illustrated in Fig.~\ref{fig:e2easr}.
\textbf{CTC} and \textbf{Transducer} are the baseline models trained based on
$\mathcal{L}_{\mathsf{ctc}}$ and $\mathcal{L}_{\mathsf{tra}}$,
as defined by Eqs.~\eqref{eq:L_ctc} and~\eqref{eq:L_tra}, respectively.
\textbf{BERT-CTC} and \textbf{BECTRA} are the proposed models trained based on
$\mathcal{L}_{\mathsf{bertctc}}$ and $\mathcal{L}_{\mathsf{bectra}}$,
as defined by Eqs.~\eqref{eq:L_bec} and~\eqref{eq:L_bectra}, respectively.

The Conformer encoder in Eq.~\eqref{eq:conformer} consisted of two convolutional neural network (CNN) layers
followed by a stack of $I=12$ encoder blocks.
The CNN layers had $256$ channels, a kernel size of $3 \times 3$, and a stride size of $2$,
which resulted in down-sampling the input length by a factor of 4 (i.e., $T=T'/4$).
For the self-attention module in Eq.~\eqref{eq:conformer_sa},
the number of heads $d_{\mathsf{h}}$, dimension of a self-attention layer $d_{\mathsf{model}}$, and
dimension of a feed-forward network $d_{\mathsf{ff}}$, were set to $4$, $256$, and $1024$, respectively.
For the depthwise separable convolution module in Eq.~\eqref{eq:conformer_conv}, we used a kernel size of $31$.

For the transducer-based models, including Transducer and BECTRA,
the prediction network was a single long short-term memory (LSTM) layer with $256$ hidden units.
The joint network consisted of a single linear layer with $256$ hidden units,
followed by a hyperbolic tangent activation function.

Regarding the BERT-based models, including BERT-CTC and BECTRA,
the concatenation network was the Transformer encoder~\cite{vaswani2017attention} with $6$ blocks,
where $d_{\mathsf{h}}$, $d_{\mathsf{model}}$, and $d_{\mathsf{ff}}$ were configured to $4$, $256$, and $2048$, respectively.
We applied two CNN layers to the encoder output before passing it through the concatenation network,
using the same configuration as that of the down-sampling layer in the Conformer encoder.
Unless stated otherwise,
we used a BERT$_{\textsc{BASE}}$ (uncased) model trained for each language from the HuggingFace Transformers library~\cite{transformers2020wolf}:
English~\cite{bert_base_uncased} (with $|\mathcal{V}^{\mathsf{b}}| = 30522$) and
Mandarin~\cite{bert_base_chinese} (with $|\mathcal{V}^{\mathsf{b}}| = 21128$).

\subsection{Training Configuration}
\label{ssec:training_configuration}
For each dataset, we mostly adhered to the configurations provided by the ESPnet2 recipe.
We trained the models for
100 epochs on LS-100, LL-10, and AS1;
70 epochs on TED2 and LS-960; and
50 epochs on CV2.
We used the Adam optimizer~\cite{kingma2015adam} for weight updates
with the beta coefficients $(\beta_1, \beta_2)$, epsilon parameter, and weight decay rate of
$(0.9, 0.999)$, $10^{-8}$, and $10^{-6}$, respectively.
We used Noam learning rate scheduling~\cite{vaswani2017attention},
where the number of warmup steps was set to $15\text{k}$, and a peak learning rate was tuned from $\{1.0, 2.0\}\times10^{-3}$.
We set the batch size to $256$, except for LL-10, which was set to $32$.
We augmented speech data using speed perturbation~\cite{ko2015audio} with a factor of $3$ and
SpecAugment~\cite{park2019specaugment,park2020specaugment}.
For the hyperparameters in SpecAugment, we set the number of frequency and time masks to $2$ and $5$,
and the size of frequency and time masks to $27$ and $0.05T'$.
For BECTRA, we set $\lambda$ in Eq.~\eqref{eq:L_bectra} to $0.5$.

For all of the models,
we applied the intermediate CTC regularization technique~\cite{tjandra2020deja,lee2021intermediate} to the Conformer encoder,
which has been demonstrated to enhance ASR performance.
Similarly to the CTC loss in Eq.~\eqref{eq:L_ctc}, an auxiliary CTC loss was calculated using
the output of the $6$-th encoder block $H^{(i=6)}$.
The intermediate CTC loss was based on a target sequence $W^{\mathsf{a}}$ tokenized by the ASR vocabulary $\mathcal{V}^{\mathsf{a}}$.
This is especially effective for model training with the large BERT vocabulary $\mathcal{V}^{\mathsf{b}}$,
facilitating the prediction of sparse word-level tokens in a hierarchical multi-tasking manner~\cite{fernandez2007sequence,sanabria2018hierarchical,krishna2018hierarchical,higuchi2022hierarchical}.

\subsection{Decoding Configuration}
\label{ssec:decoding_configuration}
A final model was obtained for evaluation  by averaging model parameters over $10$ checkpoints with the best validation performance.
For the number of iterations in BERT-CTC decoding (in Algorithm~\ref{algo:bertctc_inference}),
we set $K$ to $20$ for BERT-CTC and $10$ for BECTRA.
We performed the beam search decoding with a beam size of $10$ for Transducer and $5$ for BECTRA.

During the initialization process of a masked sequence in BERT-CTC decoding (Algorithm~\ref{algo:bertctc_inference} Step \texttt{1}),
the output length was determined based on intermediate predictions.
To be more specific, we utilized the intermediate encoder states, where the auxiliary CTC loss was applied, to perform best path decoding.
This allowed us to generate a sequence tokenized based on the ASR vocabulary $\mathcal{V}^{\mathsf{a}}$,
which was then retokenized using the BERT vocabulary $\mathcal{V}^{\mathsf{b}}$ to estimate the initial length.

\section{Results}
\label{sec:results}

\subsection{Effectiveness of BERT-CTC}
\label{ssec:results_bertctc}
\begin{figure}[t]
\centering
\resizebox{0.98\columnwidth}{!}{
\begin{tikzpicture}
\definecolor{clr1}{RGB}{31, 119, 180}
\definecolor{clr2}{RGB}{255, 127, 14}
\definecolor{clr3}{RGB}{44, 160, 44}

\pgfplotsset{
  scale only axis,
}

\begin{axis}[
    height=4.5cm,
    width=9cm,
    grid=major,
    axis y line*=left,
    xlabel=\# Iterations $K$,
    ylabel=Dev WER $\text{[\%]}$ ($\leftarrow$),
    xlabel style={font=\normalsize},
    ylabel style={font=\normalsize},
    legend cell align={left},
    ymin=7,
    ymax=15,
    ytick={8,...,14},
    yticklabels={8,,10,,12,,14},
]

\addplot[mark=*, clr1, very thick]
  coordinates{
    (1, 14.05)
    (5, 12.3)
    (10, 11.85)
    (15, 11.7)
    (20, 11.65)
    (25, 11.65)
    (30, 11.65)
  };
\label{plot_1_y1}

\addplot[mark=triangle*, mark size=2.5pt, clr2, very thick]
  coordinates{
    (1, 8.9)
    (5, 8.5)
    (10, 8.2)
    (15, 8.1)
    (20, 8.1)
    (25, 8.1)
    (30, 8.2)
  }; \label{plot_2_y1}

\end{axis}

\begin{axis}[
    height=4.5cm,
    width=9cm,
    axis y line*=right,
    axis x line=none,
    ylabel=Dev CER $\text{[\%]}$ ($\leftarrow$),
    xlabel style={font=\normalsize},
    ylabel style={font=\normalsize},
    ymin=3.333333,
    ymax=6.0,
    ytick={3,...,5},
    yticklabels={,4,5},
    legend cell align={left},
]
\addlegendimage{/pgfplots/refstyle=plot_1_y1}\addlegendentry{\ LibriSpeech-100h (WER)}
\addlegendimage{/pgfplots/refstyle=plot_2_y1}\addlegendentry{\ TED-LIUM2 (WER)}

\addplot[mark=square*, clr3, very thick]
  coordinates{
    (1, 5.3)
    (5, 5.0)
    (10, 4.1)
    (15, 3.9)
    (20, 3.9)
    (25, 3.9)
    (30, 3.9)
  }; \label{plot_1_y2}

\addlegendimage{/pgfplots/refstyle=plot_1_y2}\addlegendentry{\ AISHELL-1 (CER)}

\end{axis}

\end{tikzpicture}
}
\vspace{-0.3cm}
\caption{WER or CER [\%] ($\downarrow$) of BERT-CTC evaluated on development sets, using varying numbers of decoding iterations. When $K=1$, the model relies solely on audio information to predict output tokens. When $K>1$, the model incorporates linguistic information from BERT to refine its outputs.}
\label{fig:iterations}
\end{figure}
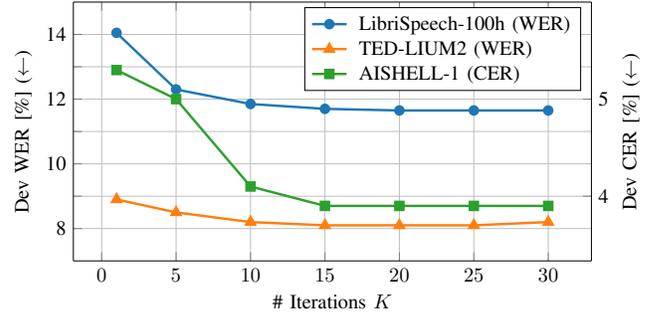

We first investigate the effectiveness of BERT-CTC,
which is designed to enhance ASR performance by integrating audio information with linguistic knowledge from BERT.
Figure~\ref{fig:iterations} depicts the relationship between
the number of decoding iterations ($K$ in Algorithm~\ref{algo:bertctc_inference}) and the BERT-CTC results,
as evaluated by the word error rate (WER) for LS-100 and TED2, and the character error rate (CER) for AS-1.
Note that WERs for LS-100 are calculated as an average of the scores on the \textit{dev-}\{\textit{clean, other}\} sets.
During decoding with $K=1$, the model only relies on the speech input,
as all output tokens are masked, and the model does not have access to any linguistic cues (cf. Eq.~\eqref{eq:tilde_W_approx}).
By increasing the number of iterations with $K>1$,
the model successfully utilized the knowledge from BERT to refine the output tokens (cf. Eq.~\eqref{eq:bertctc_inference_approx}),
leading to more beneficial outcomes across all tasks.

\subsection{Difficulty of Training ASR with BERT Vocabulary}
\label{ssec:results_vocab}
\begin{table}[t]
    \centering
    \caption{WER [\%] ($\downarrow$) of CTC and Transducer baselines compared to our proposed BERT-CTC, evaluated on LibriSpeech-100h. CTC and Transducer were trained using either the ASR vocabulary $\mathcal{V}^{\mathsf{a}}$ or the BERT vocabulary $\mathcal{V}^{\mathsf{b}}$.}
    \label{tb:vocab}
    \begin{tabular}{lx{0.7cm}x{0.7cm}x{0.6cm}x{0.6cm}x{0.6cm}x{0.6cm}}
        \toprule
        & \multicolumn{2}{c}{\textbf{Output Vocab.}} & \multicolumn{2}{c}{\textbf{Dev WER}} & \multicolumn{2}{c}{\textbf{Test WER}} \\
        \cmidrule(l{0.3em}r{0.3em}){2-3}\cmidrule(l{0.3em}r{0.3em}){4-5}\cmidrule(l{0.3em}r{0.3em}){6-7}
        \textbf{Model} & $\mathcal{V}^{\mathsf{a}}$ & $\mathcal{V}^{\mathsf{b}}$ & clean & other & clean & other \\
        \cmidrule{1-7}
        \multirow{2}{*}[0pt]{CTC} & \checkmark & & \pz6.9 & 20.1 & \pz7.0 & 20.2 \\
        & & \checkmark & 11.2 & 21.4 & 11.4 & 22.0 \\
        \cmidrule{1-7}
        \multirow{2}{*}[0pt]{Transducer} & \checkmark & & \pz5.9 & 17.7 & \pz6.0 & 17.6 \\
        & & \checkmark & \pz9.7 & 21.5 & \pz9.8 & 22.3 \\
        \cmidrule{1-7}
        BERT-CTC & & \checkmark & \pz7.0 & 16.4 & \pz7.1 & 16.5 \\
        \bottomrule
    \end{tabular}
\end{table}
In Table~\ref{tb:vocab},
we compare WERs obtained by training CTC and Transducer models on LS-100
using either the ASR or BERT vocabulary ($\mathcal{V}^{\mathsf{a}}$ vs.\ $\mathcal{V}^{\mathsf{b}}$).
It is apparent that CTC and Transducer produced notably higher WERs when trained with the BERT vocabulary,
indicating that employing word-level BERT units was not an optimal choice for ASR training~\cite{soltau2016neural}.
Moreover, we also mention that the BERT vocabulary is not precisely aligned with the intended domain
of the LibriSpeech task (e.g., Wikipedia vs.\ audiobook).
As a result, there was a potential domain mismatch between the ASR training text and BERT vocabulary.

In contrast, despite using the same BERT vocabulary,
our proposed BERT-CTC model significantly outperformed both the conventional CTC and Transducer models.
The gain from CTC demonstrates the effectiveness of leveraging BERT's contextualized linguistic embeddings
for relaxing the conditional independence assumption (cf. Table~\ref{tb:formulations}).
BERT-CTC improved over Transducer by modeling output dependencies using BERT,
allowing the model to consider bi-directional context in the target sequence (cf. Table~\ref{tb:formulations}).
BERT-CTC addressed the domain-mismatch issue through the effective use of powerful representations obtained from BERT.
However, the effectiveness of BERT-CTC diminishes when compared to CTC and Transducer models trained on the ASR vocabulary, 
thereby reducing the advantage gained from using BERT.

\subsection{Main Results}
\label{ssec:results_main}
\begin{table*}[t]
    \centering
    \caption{WER or CER [\%] ($\downarrow$) of our propoced BERT-CTC and BECTRA compared to Transducer baseline, evaluated on major ASR tasks}
    \label{tb:main_results}
    \begin{tabular}{lx{0.7cm}x{0.7cm}x{0.6cm}x{0.6cm}x{0.6cm}x{0.6cm}x{0.6cm}x{0.6cm}x{0.6cm}x{0.6cm}x{0.6cm}x{0.6cm}x{0.6cm}x{0.6cm}}
        \toprule
        & & & \multicolumn{4}{c}{\textbf{LibriSpeech-100h}} & \multicolumn{4}{c}{\textbf{LibriSpeech-960h}} & \multicolumn{2}{c}{\textbf{TED-LIUM2}} & \multicolumn{2}{c}{\textbf{AISHELL-1}} \\
        \cmidrule(l{0.3em}r{0.3em}){4-7}\cmidrule(l{0.3em}r{0.3em}){8-11}\cmidrule(l{0.3em}r{0.3em}){12-13}\cmidrule(l{0.3em}r{0.3em}){14-15}
        & \multicolumn{2}{c}{\textbf{Output Vocab.}} & \multicolumn{2}{c}{Dev WER} & \multicolumn{2}{c}{Test WER} & \multicolumn{2}{c}{Dev WER} & \multicolumn{2}{c}{Test WER} & \multicolumn{2}{c}{WER} & \multicolumn{2}{c}{CER} \\
        \cmidrule(l{0.3em}r{0.3em}){2-3}\cmidrule(l{0.3em}r{0.3em}){4-5}\cmidrule(l{0.3em}r{0.3em}){6-7}\cmidrule(l{0.3em}r{0.3em}){8-9}\cmidrule(l{0.3em}r{0.3em}){10-11}\cmidrule(l{0.3em}r{0.3em}){12-13}\cmidrule(l{0.3em}r{0.3em}){14-15}
        \textbf{Model} & $\mathcal{V}^{\mathsf{a}}$ & $\mathcal{V}^{\mathsf{b}}$ & clean & other & clean & other & clean & other & clean & other & Dev & Test & Dev & Test \\
        \midrule
        Transducer & \checkmark & & 5.9 & 17.7 & 6.0 & 17.6 & \textbf{2.5} & 6.8 & \textbf{2.8} & 6.8 & 7.8 & 7.4 & 4.9 & 5.3 \\
        BERT-CTC & & \checkmark & 7.0 & 16.4 & 7.1 & 16.5 & 3.1 & 7.1 & 3.2 & 7.1 & 8.3 & 7.6 & 3.9 & 4.0 \\
        BECTRA & \checkmark & & \textbf{5.1} & \textbf{15.4} & \textbf{5.4} & \textbf{15.5} & 2.6 & \textbf{6.7} & 2.9 & \textbf{6.7} & \textbf{7.3} & \textbf{6.9} & \textbf{3.7} & \textbf{3.9} \\
        \bottomrule
    \end{tabular}
\end{table*}
Table~\ref{tb:main_results} lists results on the major ASR tasks, including LS-100, LS-960, TED2, and AS1,
evaluated in terms of the WER or CER.
We compare our proposed models, BERT-CTC and BECTRA, with Transducer trained on the ASR vocabulary,
which was the best performing baseline from Table~\ref{tb:vocab}.
As discussed in Section~\ref{ssec:results_vocab},
BERT-CTC underperformed compared to Transducer in several tasks,
which we attribute to the vocabulary discrepancy.
Overall, BECTRA achieved the highest performance compared to all other models,
taking the advantages of both BERT-CTC and Transducer (cf. Section~\ref{ssec:formulations}).
BECTRA effectively utilized BERT knowledge by adopting BERT-CTC-based feature extraction, and
incorporated the transducer framework to enable more suitable and flexible token generation using the ASR vocabulary.
In Section~\ref{ssec:error_analysis}, we present the specific errors that BECTRA succeeded in recovering, as compared to BERT-CTC.

Another notable observation was that with more training data in LS-960,
the performance gap between Transducer and BECTRA narrowed, and the impact of BERT became less pronounced.
This finding led us to explore two further directions for investigation,
which we discuss in the following two subsections.

\subsection{Results on Low-Resource Setting}
\begin{table}[t]
    \centering
    \caption{WER [\%] ($\downarrow$) comparison on low-resource Libri-Light-10h}
    \label{tb:librilight}
    \begin{tabular}{lx{0.7cm}x{0.7cm}x{0.6cm}x{0.6cm}x{0.6cm}x{0.6cm}}
        \toprule
        & \multicolumn{2}{c}{\textbf{Output Vocab.}} & \multicolumn{2}{c}{\textbf{Dev WER}} & \multicolumn{2}{c}{\textbf{Test WER}} \\
        \cmidrule(l{0.3em}r{0.3em}){2-3}\cmidrule(l{0.3em}r{0.3em}){4-5}\cmidrule(l{0.3em}r{0.3em}){6-7}
        \textbf{Model} & $\mathcal{V}^{\mathsf{a}}$ & $\mathcal{V}^{\mathsf{b}}$ & clean & other & clean & other\\
        \midrule
        CTC & & \checkmark & 36.2 & 46.9 & 36.8 & 47.7 \\
        Transducer & & \checkmark & 34.9 & 45.9 & 35.6 & 46.8 \\
        BERT-CTC & & \checkmark & \textbf{27.2} & \textbf{39.2} & \textbf{28.3} & \textbf{40.4} \\
        \cdashlinelr{1-7}
        CTC & \checkmark & & 24.8 & 43.2 & 25.8 & 44.4 \\
        Transducer & \checkmark & & 21.7 & 38.8 & 22.3 & 39.7 \\
        BECTRA & \checkmark & & \textbf{19.9} & \textbf{36.1} & \textbf{20.3} & \textbf{37.2} \\
        \bottomrule
    \end{tabular}
\end{table}
The proposed models, BERT-CTC and BECTRA, were less significant in the LS-960 task,
likely due to the fact that the dataset already contained a sufficient amount of text data.
Consequently, the ASR models were capable of acquiring rich linguistic information specific to the LibriSpeech domain,
without relying on BERT knowledge.
This can be consistent with the recent results on LibriSpeech,
which indicate an LM (for shallow fusion) has a limited effect on performance with a well-trained ASR model~\cite{zhang2020pushing}.

We, thus, examine the other end of the spectrum,
evaluating the proposed models on LL-10, an extremely low-resource condition with only 10 hours of training data.
Table~\ref{tb:librilight} lists WERs obtained by training the models
using either the ASR or BERT vocabulary ($\mathcal{V}^{\mathsf{a}}$ vs.\ $\mathcal{V}^{\mathsf{b}}$).
The overall trend was in line with what was observed in the previous results in Tables~\ref{tb:vocab} and~\ref{tb:main_results},
highlighting the ability of the BERT-based approaches to compensate for the limited availability of training text data.

\subsection{Results on Preserving Punctuation and Casing}
\label{ssec:punct_casing}
\begin{table}[t]
    \centering
    \caption{WER [\%] ($\downarrow$) comparison on CoVoST2 with and without considering punctuation or casing. CTC and Transducer were trained on the ASR vocabulary. $\dagger$ indicates a ``cased'' model.}
    \label{tb:covost2}
    \begin{tabular}{llx{0.9cm}x{0.9cm}x{0.65cm}x{0.65cm}}
        \toprule
        & & \multicolumn{2}{c}{\textbf{Text Style}} & \multicolumn{2}{c}{\textbf{WER}} \\
        \cmidrule(l{0.3em}r{0.3em}){3-4}\cmidrule(l{0.3em}r{0.3em}){5-6}
        \textbf{Model} & \textbf{Masked LM} & Punct. & Casing & Dev & Test \\
        \midrule
        CTC & -- & -- & -- & 18.7 & 23.2 \\
        Transducer & -- & -- & -- & 15.2 & 18.8 \\
        \cmidrule{1-6}
        & BERT$_\textsc{base}$ & \ding{55} & \ding{55} & 14.4 & 17.6 \\
        & BERT$_\textsc{base}$ & \checkmark & \ding{55} & \textbf{14.0} & 17.2 \\
        & BERT$_\textsc{base}^{\dagger}$ & \checkmark & \checkmark & \textbf{14.0} & \textbf{17.1} \\
        & RoBERTa$_\textsc{base}^{\dagger}$ & \checkmark & \checkmark & 14.1 & \textbf{17.1} \\
        \cdashlinelr{2-6}
        BECTRA & BERT$_\textsc{large}$ & \checkmark & \ding{55} & \textbf{13.4} & \textbf{16.4} \\
        & BERT$_\textsc{large}^{\dagger}$ & \checkmark & \checkmark & 13.8 & 16.7 \\
        \cdashlinelr{2-6}
        & DistilBERT$_\textsc{base}$ & \checkmark & \ding{55} & \textbf{14.3} & \textbf{17.6} \\
        & DistilBERT$_\textsc{base}^{\dagger}$ & \checkmark & \checkmark & 14.9 & 17.9 \\
        & ALBERT$_\textsc{base}$ & \checkmark & \ding{55} & 14.6 & 17.7 \\
        \bottomrule
    \end{tabular}
\end{table}
The experimental setups used to obtain results in Table~\ref{tb:main_results} could potentially limit the full capabilities of BERT
since the training text data was normalized for ASR training (see Section~\ref{ssec:experimental_setting_data}).
Pre-trained LMs are typically trained on written-style text,
preserving punctuation and casing as a standard practice.
Therefore, it appears that the prior experimental condition may have caused a discrepancy regarding the text format used as input into BERT.

To verify the above consideration, we conducted experiments on CV2
while explicitly controlling the preservation of punctuation and casing.
Table~\ref{tb:covost2} presents the results of BECTRA
in comparison to CTC and Transducer trained on the ASR vocabulary.
Note that punctuation and casing only affect the BERT-CTC processing (i.e., Algorithm~\ref{algo:bectra_inference} Step \texttt{1})
with the BERT vocabulary $\mathcal{V}^{\mathsf{b}}$, and
the WER is calculated using the normalized text,
which is obtained from a hypothesis tokenized by the ASR vocabulary $\mathcal{V}^{\mathsf{a}}$.
For training with casing preserved, we used the ``cased'' model\footnote{\url{https://huggingface.co/bert-base-cased}}.
Looking at the BECTRA results based on BERT$_{\textsc{base}}$,
BECTRA outperformed the baseline models even without considering punctuation and casing,
which is consistent with the outcome in Table~\ref{tb:main_results}.
The addition of punctuation provided further improvement,
whereas the impact of adding casing was less significant.
This suggests the importance of matching the input text format used for BERT
with that used for input during pre-training.

BECTRA is capable of processing two distinct types of text with different vocabularies,
thanks to the BERT-CTC's ability to handle BERT information and
the transducer decoder's capacity to generate tokens in a desired text style.
This is accomplished through an end-to-end framework,
eliminating the need for a retokenization process in the output sequence.

\subsection{Application of BERT Variants}
In Table~\ref{tb:covost2},
we compare the BECTRA results obtained from using various pre-trained masked LMs
other than BERT$_{\textsc{base}}$.
RoBERTa$_\textsc{base}$\footnote{\url{https://huggingface.co/roberta-base}} is an extension of BERT that is
constructed with an improved pre-training procedure~\cite{liu2019roberta}.
However, there was little improvement over the results using vanilla BERT$_\textsc{base}$.
BECTRA greatly benefited from increasing the capacity of a pre-trained LM,
with BERT$_{\textsc{large}}$\footnote{\url{https://huggingface.co/bert-large-{cased,uncased}}} achieving the best overall performance.

BECTRA incurs a high computational cost, especially during inference,
primarily due to the multiple forward passes in BERT (i.e., $K=10$ times)
with the $\mathcal{O}(N^2)$ computational and memory complexities in self-attention layers.
To mitigate this drawback, we explored lightweight variants,
including DistilBERT$_\textsc{base}$\footnote{\url{https://huggingface.co/distilbert-base-{cased,uncased}}}
and ALBERT$_\textsc{base}$\footnote{\url{https://huggingface.co/albert-base-v2}}.
DistilBERT distills BERT's knowledge into a more compact model~\cite{sanh2019distilbert}, while
ALBERT reduces model size by sharing common parameters across layers~\cite{lan2020albert}.
Both lightweight models achieved superior results compared to the baseline models,
with only minor performance degradation compared to BERT$_\textsc{base}$.

In alignment with the observation in Section~\ref{ssec:punct_casing},
the BERT variants gave more importance to considering punctuation than casing.

\subsection{Combination with Shallow Fusion}
\begin{table}[t]
    \centering
    \caption{WER [\%] ($\downarrow$) comparison on LibriSpeech-100h with and without performing shallow fusion during inference}
    \label{tb:shallow_fusion}
    \tabcolsep 4pt
    \begin{tabular}{lcccccccc}
        \toprule
        & \multicolumn{4}{c}{w/o Shallow Fusion} & \multicolumn{4}{c}{w/ Shallow Fusion} \\
        \cmidrule(l{0.3em}r{0.3em}){2-5}\cmidrule(l{0.3em}r{0.3em}){6-9}
        & \multicolumn{2}{c}{\textbf{Dev WER}} & \multicolumn{2}{c}{\textbf{Test WER}} & \multicolumn{2}{c}{\textbf{Dev WER}} & \multicolumn{2}{c}{\textbf{Test WER}} \\
        \cmidrule(l{0.3em}r{0.3em}){2-3}\cmidrule(l{0.3em}r{0.3em}){4-5}\cmidrule(l{0.3em}r{0.3em}){6-7}\cmidrule(l{0.3em}r{0.3em}){8-9}
        \textbf{Model} & clean & other & clean & other & clean & other & clean & other \\
        \midrule
        Transducer & 5.9 & 17.7 & 6.0 & 17.6 & 5.1 & 15.0 & 5.1 & 15.1 \\
        BECTRA & \textbf{5.1} & \textbf{15.4} & \textbf{5.4} & \textbf{15.5} & \textbf{4.5} & \textbf{14.2} & \textbf{4.9} & \textbf{14.2} \\
        \bottomrule
    \end{tabular}
\end{table}
We examined the feasibility of utilizing an in-domain LM during BECTRA inference.
BECTRA can adopt the commonly used shallow fusion technique,
as its inference process relies on the original transducer framework (Algorithm~\ref{algo:bectra_inference} Step \texttt{2}).
We used the external text data provided by LibriSpeech to train an RNN-LM, which consisted of $4$ LSTM layers with $2048$ units.
The LM weight and beam size for shallow fusion were configured to $0.5$ and $20$, respectively.

In Table~\ref{tb:shallow_fusion}, we compare the WER between the Transducer and BECTRA models,
which were trained on LS-100 and decoded with or without the RNN-LM.
Through the incorporation of linguistic knowledge from RNN-LM via shallow fusion,
Transducer significantly improved the performance,
resulting in lower WERs compared to BECTRA, which by default employs BERT in its formulation.
Similarly, the performance of BECTRA was further improved by utilizing shallow fusion.
This indicates that BECTRA effectively integrated general knowledge from BERT and domain-specific knowledge from the RNN-LM,
thereby enhancing its ability to consider linguistic information.

\section{Analysis}
\label{sec:analyses}

\subsection{Example Decoding Process of BECTRA}
\label{ssec:error_analysis}
\begin{table*}[t]
    \centering
    \caption{Example inference process of BECTRA (Algorithm~\ref{algo:bectra_inference}), decoding an utterance from CoVoST2 test set. During BERT-CTC inference (Algorithm~\ref{algo:bectra_inference} Step \texttt{1}), the highlighted tokens were replaced with the mask token and subsequently re-predicted in the following iteration. The transducer decoding with beam search further refined the BERT-CTC result (Algorithm~\ref{algo:bectra_inference} Step \texttt{2}). The corrected tokens are colored in blue, while the incorrect ones are in red. Punctuation is only considered during BERT-CTC decoding.}
    \label{tb:decoding_example}
    \begin{tabular}{c@{\hspace{0.15cm}}ll}
        \toprule
        \multirow{4}{*}[0pt]{\begin{rotatebox}{90}{$\underset{\xleftarrow{\hspace*{1.1cm}}}{\text{Refine}}$}\end{rotatebox}} &
        BERT-CTC ($k\!=\!1$) & \bgmask{car} \bgmask{'} \bgmask{s} \bgmask{business} \bgmask{in} \bgmask{mc} \bgmask{cly} \bgmask{was} \bgmask{in} \bgmask{sawmill} \bgmask{mills} \bgmask{,} \bgmask{tutine} \bgmask{,} \bgmask{lumber} \bgmask{,} \bgunmask{and} \bgmask{land} \bgmask{.} \\
        & BERT-CTC ($k\!=\!5$) & \bgmask{carr} \bgunmask{'} \bgunmask{s} \bgunmask{business} \bgmask{in} \bgmask{mclean} \bgunmask{was} \bgunmask{in} \bgmask{sawmill} \bgmask{mills} \bgmask{,} \bgmask{tu} \bgunmask{,} \bgunmask{lumber} \bgunmask{and} \bgunmask{land} \bgunmask{.} \\
        & BERT-CTC ($k\!=\!10$) & \bgunmask{\blue{carr}} \bgunmask{'} \bgunmask{s} \bgunmask{business} \bgunmask{in} \bgunmask{\red{mclean}} \bgunmask{was} \bgunmask{in} \bgunmask{\blue{sawmills}} \bgunmask{,} \bgunmask{\red{carpenter}} \bgunmask{,} \bgunmask{lumber} \bgunmask{and} \bgunmask{land} \bgunmask{.} \\
        & BECTRA ($B\!=\!5$) & \bgunmask{carr} \bgunmask{\&apos} \bgunmask{s} \bgunmask{business} \bgunmask{in} \bgunmask{\blue{mcclenny}} \bgunmask{was} \bgunmask{in} \bgunmask{sawmills} \bgunmask{\blue{turpentine}} \bgunmask{lumber} \bgunmask{and} \bgunmask{land} \\
        \cdashlinelr{1-3}
        & Reference & \bgunmask{carr} \bgunmask{\&apos} \bgunmask{s} \bgunmask{business} \bgunmask{in} \bgunmask{mcclenny} \bgunmask{was} \bgunmask{in} \bgunmask{sawmills} \bgunmask{turpentine} \bgunmask{lumber} \bgunmask{and} \bgunmask{land} \\
        \bottomrule
    \end{tabular}
\end{table*}
Table~\ref{tb:decoding_example} provides an example of the BECTRA inference process,
which was obtained by decoding an utterance in the CoVoST2 test set.
We used the model trained with BERT$_\textsc{base}$ from Table~\ref{tb:covost2},
which only preserved punctuation during BERT-CTC decoding.
During the first iteration of BERT-CTC inference ($k=1$),
the model produced erroneous predictions that are phonetically similar to the actual tokens
(e.g., ``car'' vs.\ ``carr'', ``mc cly'' vs.\ ``mcclenny'').
The model was solely conditioned on acoustic information during the first iteration,
which led to difficulties in accurately determining the target tokens.
As the iteration progressed ($k=\{5,10\}$), the model was able to correct some of the errors by
considering the output dependencies extracted from BERT.
However, the model still struggled with recognizing rare words such as the place name ``mcclenny''.
In addition, the model mistakenly identified the technical term ``turpentine'' as ``carpenter'',
despite the two words sounding dissimilar.
This error is likely due to the contextual information being influenced by the BERT knowledge.
The transducer decoding in BECTRA effectively recovered these errors by accurately predicting the rare words.
The autoregressive token generation facilitated more flexible estimation of tokens
using a vocabulary suited for ASR in the CoVoST2 domain.

\subsection{Attention Visualization}
\label{ssec:attention_visualization}
\begin{figure}[t]
    \centering
    \includegraphics[width=0.90\linewidth]{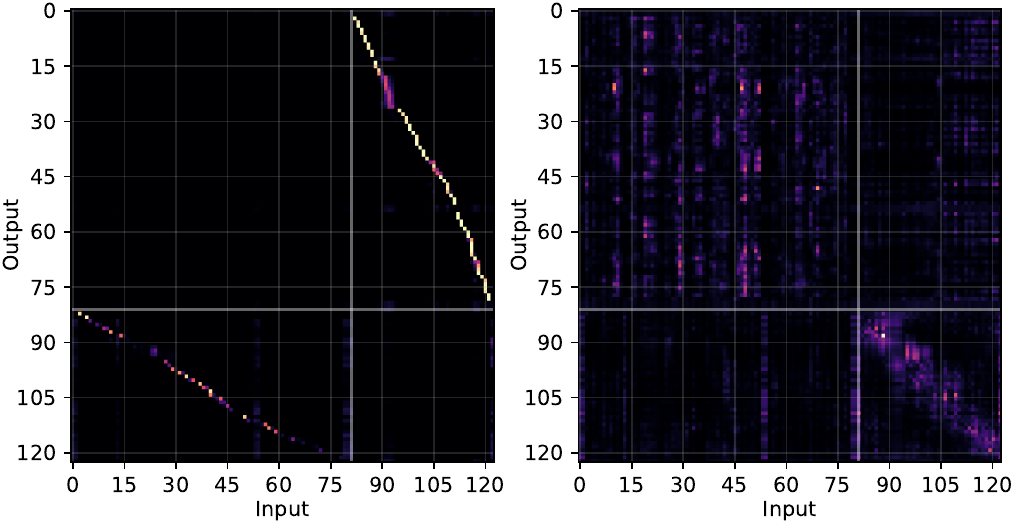}
    \caption{Visualization of self-attention weights learned in the concatenation network. White lines indicate the boundaries of audio and token sequences, $H$ and $E$, which are concatenated and processed by self-attention in Eq.~\eqref{eq:p_bec_a_tW_O}.}
    \label{fig:attention}
\end{figure}
In Fig.~\ref{fig:attention}, we present example attention weight matrices that were
obtained from the second self-attention layer of the concatenation network in BERT-CTC.
We identified two major attention patterns:
weights aligning audio and token sequences by capturing their inter-dependencies (Fig.~\ref{fig:attention} left) and
weights attending to the inner-dependencies within each sequence (Fig.~\ref{fig:attention} right).
These attention weights support the effectiveness of our proposed architectural design for BERT-CTC,
indicating audio and linguistic information are merged by considering their inter end inner-dependencies.

\subsection{Conditional Independence of $p(W^{\mathsf{b}}|\tilde{W}^{\mathsf{b}}, \cancel{O})$}
\label{ssec:adapter_bert}
We empirically validate the conditional independence assumption made in Eq.~\eqref{eq:p_W_tW_O_approx},
where the output sequence $W^{\mathsf{b}}$ depends solely on its masked sequence $\tilde{W}^{\mathsf{b}}$ without audio information $O$.
To this end, we incorporated trainable cross-attention layers into the BERT module,
which is similar to the technique proposed in Adapter-BERT Networks~\cite{guo2020incorporating}. 
These additional layers enable each BERT layer to attend to the audio encoder output $H$,
thereby allowing BERT-CTC to achieve $p(W^{\mathsf{b}}|\tilde{W}^{\mathsf{b}},O)$.
After training the modieifed BERT-CTC on LS-100,
we observed inferior WERs compared to the original BERT-CTC presented in Table~\ref{tb:main_results},
with 7.2\%/17.9\% and 7.3\%/18.0\% on the development and test sets, respectively.

The finding above suggests that BERT is capable of capturing sophisticated linguistic information
without relying on audio input conditioning.
Furthermore, this implies that our proposed formulation does not require
any adaptation or fine-tuning of BERT.

\subsection{BECTRA with BERT Vocabulary}
\label{ssec:bectra_bert}
To further examine the advantages of BECTRA compared to BERT-CTC,
we utilized the BERT vocabulary to train the transducer decoder of BECTRA.
Under the same experimental conditions using LS-100,
BECTRA with the BERT vocabulary achieved 7.1\%/16.6\% on the LibriSpeech test sets,
while BERT-CTC resulted in slightly worse scores of 7.3\%/16.9\%.
This improvement over BERT-CTC can be attributed to the transducer-based formulation in BECTRA,
which does not rely on the conditional independence assumption between outputs
(Eq.~\eqref{eq:p_bec_A_W_O_approx} vs.\ Eq.\eqref{eq:p_bectra_W_tW_O}).
Nevertheless, using the ASR vocabulary appeared to be the superior choice.

Different from the results reported in Table~\ref{tb:main_results},
the above results were obtained by reducing the number of training epochs ($100\rightarrow50$).
Training a transducer-based model with a large vocabulary size leads to a substantial increase in memory consumption~\cite{lee2022memory},
resulting in a significant extension of the training time.
Therefore, employing the ASR vocabulary is the optimal approach for constructing the BECTRA model,
as it also facilitates faster training and inference.

\section{Conclusion and Future Works}
We proposed novel end-to-end ASR models that effectively incorporate a pre-trained masked LM in their formulations.
BERT-CTC adapts BERT for CTC to alleviate the conditional independence assumption between output tokens,
while integrating both audio and linguistic information through an iterative refinement algorithm.
BECTRA extends BERT-CTC by adopting the transducer framework,
which enables the management of different vocabularies that contain varying text formats and styles.
The experimental results conducted on diverse datasets
demonstrated that the advantages of the proposed models outperform the conventional approaches.
Furthermore, our comprehensive analyses confirmed the effectiveness of the proposed formulations and architectural designs.

To further expand the scope of this study,
future work should consider the implementation of the proposed models in streaming scenarios.
The possible solutions include the adoption of the two-pass modeling~\cite{sainath2019two}
or blockwise-attention mechanism~\cite{wang2021streaming}.
In addition, we plan to extend our approaches to incorporate autoregressive pre-trained LMs,
such as GPT-3~\cite{brown2020language} and LLaMA~\cite{touvron2023llama}.

\bibliographystyle{IEEEtran}
\bibliography{refs_short}

\begin{thebibliography}{10}
\providecommand{\url}[1]{#1}
\csname url@samestyle\endcsname
\providecommand{\newblock}{\relax}
\providecommand{\bibinfo}[2]{#2}
\providecommand{\BIBentrySTDinterwordspacing}{\spaceskip=0pt\relax}
\providecommand{\BIBentryALTinterwordstretchfactor}{4}
\providecommand{\BIBentryALTinterwordspacing}{\spaceskip=\fontdimen2\font plus
\BIBentryALTinterwordstretchfactor\fontdimen3\font minus
  \fontdimen4\font\relax}
\providecommand{\BIBforeignlanguage}[2]{{%
\expandafter\ifx\csname l@#1\endcsname\relax
\typeout{** WARNING: IEEEtran.bst: No hyphenation pattern has been}%
\typeout{** loaded for the language `#1'. Using the pattern for}%
\typeout{** the default language instead.}%
\else
\language=\csname l@#1\endcsname
\fi
#2}}
\providecommand{\BIBdecl}{\relax}
\BIBdecl

\bibitem{devlin2019bert}
J.~Devlin \emph{et~al.}, ``{BERT}: Pre-training of deep bidirectional
  {Transformers} for language understanding,'' in \emph{Proc. NAACL-HLT}, 2019,
  pp. 4171--4186.

\bibitem{brown2020language}
T.~Brown \emph{et~al.}, ``Language models are few-shot learners,'' in
  \emph{Proc. NeurIPS}, 2020, pp. 1877--1901.

\bibitem{tenney2019bert}
I.~Tenney \emph{et~al.}, ``{BERT} rediscovers the classical {NLP} pipeline,''
  in \emph{Proc. ACL}, 2019, pp. 4593--4601.

\bibitem{shin2019effective}
J.~Shin \emph{et~al.}, ``Effective sentence scoring method using {BERT} for
  speech recognition,'' in \emph{Proc. ACML}, 2019, pp. 1081--1093.

\bibitem{huang2021speech}
W.-C. Huang \emph{et~al.}, ``Speech recognition by simply fine-tuning {BERT},''
  in \emph{Proc. ICASSP}, 2021, pp. 7343--7347.

\bibitem{chuang2020speechbert}
Y.-S. Chuang \emph{et~al.}, ``{SpeechBERT}: An audio-and-text jointly learned
  language model for end-to-end spoken question answering,'' in \emph{Proc.
  Interspeech}, 2020, pp. 4168--4172.

\bibitem{chung2021splat}
Y.-A. Chung \emph{et~al.}, ``{SPLAT}: Speech-language joint pre-training for
  spoken language understanding,'' in \emph{Proc. NAACL-HLT}, 2021, pp.
  1897--1907.

\bibitem{hayashi2019pre}
T.~Hayashi \emph{et~al.}, ``Pre-trained text embeddings for enhanced
  text-to-speech synthesis,'' in \emph{Proc. Interspeech}, 2019, pp.
  4430--4434.

\bibitem{kenter2020improving}
T.~Kenter \emph{et~al.}, ``Improving the prosody of {RNN}-based english
  text-to-speech synthesis by incorporating a {BERT} model,'' in \emph{Proc.
  Interspeech}, 2020, pp. 4412--4416.

\bibitem{bang2022improving}
J.-U. Bang \emph{et~al.}, ``Improving end-to-end speech translation model with
  {BERT}-based contextual information,'' in \emph{Proc. ICASSP}, 2022, pp.
  6227--6231.

\bibitem{graves2014towards}
A.~Graves and N.~Jaitly, ``Towards end-to-end speech recognition with recurrent
  neural networks,'' in \emph{Proc. ICML}, 2014, pp. 1764--1772.

\bibitem{chorowski2015attention}
J.~K. Chorowski \emph{et~al.}, ``Attention-based models for speech
  recognition,'' in \emph{Proc. NeurIPS}, 2015, pp. 577--585.

\bibitem{chan2016listen}
W.~Chan \emph{et~al.}, ``Listen, attend and spell: A neural network for large
  vocabulary conversational speech recognition,'' in \emph{Proc. ICASSP}, 2016,
  pp. 4960--4964.

\bibitem{futami2020distilling}
H.~Futami \emph{et~al.}, ``Distilling the knowledge of {BERT} for
  sequence-to-sequence {ASR},'' in \emph{Proc. Interspeech}, 2020, pp.
  3635--3639.

\bibitem{bai2021fast}
Y.~Bai \emph{et~al.}, ``Fast end-to-end speech recognition via
  non-autoregressive models and cross-modal knowledge transferring from
  {BERT},'' \emph{IEEE/ACM Transactions on Audio, Speech, and Language
  Processing}, vol.~29, pp. 1897--1911, 2021.

\bibitem{kubo2022knowledge}
Y.~Kubo \emph{et~al.}, ``Knowledge transfer from large-scale pretrained
  language models to end-to-end speech recognizers,'' in \emph{Proc. ICASSP},
  2022, pp. 8512--8516.

\bibitem{lu2022context}
K.-H. Lu and K.-Y. Chen, ``A context-aware knowledge transferring strategy for
  {CTC}-based {ASR},'' in \emph{Proc. SLT}, 2022, pp. 60--67.

\bibitem{salazar2020masked}
J.~Salazar \emph{et~al.}, ``Masked language model scoring,'' in \emph{Proc.
  ACL}, 2020, pp. 2699--2712.

\bibitem{chiu2021innovative}
S.-H. Chiu and B.~Chen, ``Innovative {BERT}-based reranking language models for
  speech recognition,'' in \emph{Proc. SLT}, 2021, pp. 266--271.

\bibitem{futami2021asr}
H.~Futami \emph{et~al.}, ``{ASR} rescoring and confidence estimation with
  {ELECTRA},'' in \emph{Proc. ASRU}, 2021, pp. 380--387.

\bibitem{udagawa2022effect}
T.~Udagawa \emph{et~al.}, ``Effect and analysis of large-scale language model
  rescoring on competitive {ASR} systems,'' in \emph{Proc. Interspeech}, 2022,
  pp. 3919--3923.

\bibitem{yi2021efficiently}
C.~Yi \emph{et~al.}, ``Efficiently fusing pretrained acoustic and linguistic
  encoders for low-resource speech recognition,'' \emph{IEEE Signal Processing
  Letters}, vol.~28, pp. 788--792, 2021.

\bibitem{zheng2021wav}
G.~Zheng \emph{et~al.}, ``Wav-{BERT}: Cooperative acoustic and linguistic
  representation learning for low-resource speech recognition,'' in \emph{Proc.
  Findings of EMNLP}, 2021, pp. 2765--2777.

\bibitem{deng2021improving}
K.~Deng \emph{et~al.}, ``Improving hybrid {CTC}/attention end-to-end speech
  recognition with pretrained acoustic and language models,'' in \emph{Proc.
  ASRU}, 2021, pp. 76--82.

\bibitem{yu2022non}
F.-H. Yu \emph{et~al.}, ``Non-autoregressive {ASR} modeling using pre-trained
  language models for {Chinese} speech recognition,'' \emph{IEEE/ACM
  Transactions on Audio, Speech, and Language Processing}, vol.~30, pp.
  1474--1482, 2022.

\bibitem{graves2006connectionist}
A.~Graves \emph{et~al.}, ``Connectionist temporal classification: Labelling
  unsegmented sequence data with recurrent neural networks,'' in \emph{Proc.
  ICML}, 2006, pp. 369--376.

\bibitem{graves2012sequence}
A.~Graves, ``Sequence transduction with recurrent neural networks,'' in
  \emph{Proc. ICML Representation Learning Workshop}, 2012.

\bibitem{higuchi2022bert}
Y.~Higuchi \emph{et~al.}, ``{BERT} meets {CTC}: New formulation of end-to-end
  speech recognition with pre-trained masked language model,'' in \emph{Proc.
  Findings of EMNLP}, 2022, pp. 5486--5503.

\bibitem{higuchi2023bectra}
------, ``{BECTRA}: Transducer-based end-to-end {ASR} with {BERT}-enhanced
  encoder,'' in \emph{Proc. ICASSP}, 2023.

\bibitem{graves2013speech}
A.~Graves \emph{et~al.}, ``Speech recognition with deep recurrent neural
  networks,'' in \emph{Proc. ICASSP}, 2013, pp. 6645--6649.

\bibitem{gulati2020conformer}
A.~Gulati \emph{et~al.}, ``Conformer: Convolution-augmented {Transformer} for
  speech recognition,'' in \emph{Proc. Interspeech}, 2020, pp. 5036--5040.

\bibitem{hori2017advances}
T.~Hori \emph{et~al.}, ``Advances in joint {CTC}-attention based end-to-end
  speech recognition with a deep {CNN} encoder and {RNN-LM},'' in \emph{Proc.
  Interspeech}, 2017, pp. 949--953.

\bibitem{vaswani2017attention}
A.~Vaswani \emph{et~al.}, ``Attention is all you need,'' in \emph{Proc.
  NeurIPS}, 2017, pp. 5998--6008.

\bibitem{higuchi2021comparative}
Y.~Higuchi \emph{et~al.}, ``A comparative study on non-autoregressive modelings
  for speech-to-text generation,'' in \emph{Proc. ASRU}, 2021, pp. 47--54.

\bibitem{gu2018non}
J.~Gu \emph{et~al.}, ``Non-autoregressive neural machine translation,'' in
  \emph{Proc. ICLR}, 2018.

\bibitem{boyer2021study}
F.~Boyer \emph{et~al.}, ``A study of {Transducer} based end-to-end {ASR} with
  {ESPnet}: Architecture, auxiliary loss and decoding strategies,'' in
  \emph{Proc. ASRU}, 2021, pp. 16--23.

\bibitem{fujita2020insertion}
Y.~Fujita \emph{et~al.}, ``Insertion-based modeling for end-to-end automatic
  speech recognition,'' in \emph{Proc. Interspeech}, 2020, pp. 3660--3664.

\bibitem{peters2019tune}
M.~E. Peters \emph{et~al.}, ``To tune or not to tune? adapting pretrained
  representations to diverse tasks,'' in \emph{Proc. RepL4NLP}, 2019, pp.
  7--14.

\bibitem{zhu2020incorporating}
J.~Zhu \emph{et~al.}, ``Incorporating {BERT} into neural machine translation,''
  in \emph{Proc. ICLR}, 2020.

\bibitem{stappen2020cross}
L.~Stappen \emph{et~al.}, ``Cross-lingual zero-and few-shot hate speech
  detection utilising frozen transformer language models and {AXEL},''
  \emph{arXiv preprint arXiv:2004.13850}, 2020.

\bibitem{mcclelland1986trace}
J.~L. McClelland and J.~L. Elman, ``The {TRACE} model of speech perception,''
  \emph{Cognitive psychology}, vol.~18, no.~1, pp. 1--86, 1986.

\bibitem{norris1994shortlist}
D.~Norris, ``Shortlist: A connectionist model of continuous speech
  recognition,'' \emph{Cognition}, vol.~52, no.~3, pp. 189--234, 1994.

\bibitem{ghazvininejad2019mask}
M.~Ghazvininejad \emph{et~al.}, ``Mask-predict: Parallel decoding of
  conditional masked language models,'' in \emph{Proc. EMNLP-IJCNLP}, 2019, p.
  6112–6121.

\bibitem{chan2020imputer}
W.~Chan \emph{et~al.}, ``Imputer: Sequence modelling via imputation and dynamic
  programming,'' in \emph{Proc. ICML}, 2020, pp. 1403--1413.

\bibitem{higuchi2020mask}
Y.~Higuchi \emph{et~al.}, ``{Mask CTC}: Non-autoregressive end-to-end {ASR}
  with {CTC} and mask predict,'' in \emph{Proc. Interspeech}, 2020, pp.
  3655--3659.

\bibitem{higuchi2021improved}
------, ``Improved mask-{CTC} for non-autoregressive end-to-end {ASR},'' in
  \emph{Proc. ICASSP}, 2021, pp. 8363--8367.

\bibitem{chen2021non}
N.~Chen \emph{et~al.}, ``Non-autoregressive transformer for speech
  recognition,'' \emph{IEEE Signal Processing Letters}, vol.~28, pp. 121--125,
  2021.

\bibitem{futami2022non}
H.~Futami \emph{et~al.}, ``Non-autoregressive error correction for {CTC}-based
  {ASR} with phone-conditioned masked {LM},'' in \emph{Proc. Interspeech},
  2022, pp. 3889--3893.

\bibitem{fan2022acoustic}
R.~Fan \emph{et~al.}, ``Acoustic-aware non-autoregressive spell correction with
  mask sample decoding,'' \emph{arXiv preprint arXiv:2210.08665}, 2022.

\bibitem{hannun2014deep}
A.~Hannun \emph{et~al.}, ``Deep speech: Scaling up end-to-end speech
  recognition,'' \emph{arXiv preprint arXiv:1412.5567}, 2014.

\bibitem{gulcehre2015using}
C.~Gulcehre \emph{et~al.}, ``On using monolingual corpora in neural machine
  translation,'' \emph{arXiv preprint arXiv:1503.03535}, 2015.

\bibitem{chorowski2017towards}
J.~Chorowski and N.~Jaitly, ``Towards better decoding and language model
  integration in sequence to sequence models,'' in \emph{Proc. Interspeech},
  2017, pp. 523--527.

\bibitem{kannan2018analysis}
A.~Kannan \emph{et~al.}, ``An analysis of incorporating an external language
  model into a sequence-to-sequence model,'' in \emph{Proc. ICASSP}, 2018, pp.
  5824--5828.

\bibitem{sriram2018cold}
A.~Sriram \emph{et~al.}, ``Cold fusion: Training seq2seq models together with
  language models,'' in \emph{Proc. Interspeech}, 2018, pp. 387--391.

\bibitem{shan2019component}
C.~Shan \emph{et~al.}, ``Component fusion: Learning replaceable language model
  component for end-to-end speech recognition system,'' in \emph{Proc. ICASSP},
  2019, pp. 5361--5635.

\bibitem{watanabe2018espnet}
S.~Watanabe \emph{et~al.}, ``{ESPnet}: End-to-end speech processing toolkit,''
  in \emph{Proc. Interspeech}, 2018, pp. 2207--2211.

\bibitem{watanabe20212020}
------, ``The 2020 {ESPnet} update: New features, broadened applications,
  performance improvements, and future plans,'' in \emph{Proc. DSLW}, 2021, pp.
  1--6.

\bibitem{panayotov2015librispeech}
V.~Panayotov \emph{et~al.}, ``Librispeech: An {ASR} corpus based on public
  domain audio books,'' in \emph{Proc. ICASSP}, 2015, pp. 5206--5210.

\bibitem{kahn2020libri}
J.~Kahn \emph{et~al.}, ``{Libri-Light}: A benchmark for {ASR} with limited or
  no supervision,'' in \emph{Proc. ICASSP}, 2020, pp. 7669--7673.

\bibitem{rousseau2014enhancing}
A.~Rousseau \emph{et~al.}, ``Enhancing the {TED}-{LIUM} corpus with selected
  data for language modeling and more {TED} talks,'' in \emph{Proc. LREC},
  2014, pp. 3935--3939.

\bibitem{bu2017aishell}
H.~Bu \emph{et~al.}, ``{AISHELL}-1: An open-source {Mandarin} speech corpus and
  a speech recognition baseline,'' in \emph{Proc. O-COCOSDA}, 2017, pp. 1--5.

\bibitem{wang2021covost}
C.~Wang \emph{et~al.}, ``{CoVoST} 2 and massively multilingual speech
  translation,'' in \emph{Proc. Interspeech}, 2021, pp. 2247--2251.

\bibitem{ardila2020common}
R.~Ardila \emph{et~al.}, ``Common voice: A massively-multilingual speech
  corpus,'' in \emph{Proc. LREC}, 2020, pp. 4218--4222.

\bibitem{kudo2018subword}
T.~Kudo, ``Subword regularization: Improving neural network translation models
  with multiple subword candidates,'' in \emph{Proc. ACL}, 2018, pp. 66--75.

\bibitem{transformers2020wolf}
T.~Wolf \emph{et~al.}, ``Transformers: State-of-the-art natural language
  processing,'' in \emph{Proc. EMNLP: System Demonstrations}, 2020, pp. 38--45.

\bibitem{bert_base_uncased}
``bert-base-uncased,'' \url{https://huggingface.co/bert-base-uncased}, [Online;
  Accessed on March-23-2023].

\bibitem{bert_base_chinese}
``bert-base-chinese,'' \url{https://huggingface.co/bert-base-chinese}, [Online;
  Accessed on March-23-2023].

\bibitem{kingma2015adam}
D.~P. Kingma and J.~Ba, ``Adam: A method for stochastic optimization,'' in
  \emph{Proc. ICLR}, 2015.

\bibitem{ko2015audio}
T.~Ko \emph{et~al.}, ``Audio augmentation for speech recognition,'' in
  \emph{Proc. Interspeech}, 2015, pp. 3586--3589.

\bibitem{park2019specaugment}
D.~S. Park \emph{et~al.}, ``{SpecAugment}: A simple data augmentation method
  for automatic speech recognition,'' in \emph{Proc. Interspeech}, 2019, pp.
  2613--2617.

\bibitem{park2020specaugment}
------, ``{SpecAugment} on large scale datasets,'' in \emph{Proc. ICASSP},
  2020, pp. 6879--6883.

\bibitem{tjandra2020deja}
A.~Tjandra \emph{et~al.}, ``{DEJA-VU}: Double feature presentation and iterated
  loss in deep {Transformer} networks,'' in \emph{Proc. ICASSP}, 2020, pp.
  6899--6903.

\bibitem{lee2021intermediate}
J.~Lee and S.~Watanabe, ``Intermediate loss regularization for {CTC}-based
  speech recognition,'' in \emph{Proc. ICASSP}, 2021, pp. 6224--6228.

\bibitem{fernandez2007sequence}
S.~Fern\'{a}ndez \emph{et~al.}, ``Sequence labelling in structured domains with
  hierarchical recurrent neural networks,'' in \emph{Proc. IJCAI}, 2007, pp.
  774--779.

\bibitem{sanabria2018hierarchical}
R.~Sanabria and F.~Metze, ``Hierarchical multitask learning with {CTC},'' in
  \emph{Proc. SLT}, 2018, pp. 485--490.

\bibitem{krishna2018hierarchical}
K.~Krishna \emph{et~al.}, ``Hierarchical multitask learning for {CTC}-based
  speech recognition,'' \emph{arXiv preprint arXiv:1807.06234}, 2018.

\bibitem{higuchi2022hierarchical}
Y.~Higuchi \emph{et~al.}, ``Hierarchical conditional end-to-end {ASR} with
  {CTC} and multi-granular subword units,'' in \emph{Proc. ICASSP}, 2022, pp.
  7797--7801.

\bibitem{soltau2016neural}
H.~Soltau \emph{et~al.}, ``Neural speech recognizer: {A}coustic-to-word {LSTM}
  model for large vocabulary speech recognition,'' \emph{arXiv preprint
  arXiv:1610.09975}, 2016.

\bibitem{zhang2020pushing}
Y.~Zhang \emph{et~al.}, ``Pushing the limits of semi-supervised learning for
  automatic speech recognition,'' \emph{arXiv preprint arXiv:2010.10504}, 2020.

\bibitem{liu2019roberta}
Y.~Liu \emph{et~al.}, ``{RoBERTa}: A robustly optimized {BERT} pretraining
  approach,'' \emph{arXiv preprint arXiv:1907.11692}, 2019.

\bibitem{sanh2019distilbert}
V.~Sanh \emph{et~al.}, ``{DistilBERT}, a distilled version of {BERT}: smaller,
  faster, cheaper and lighter,'' in \emph{Proc. NeurIPS Workshop on Energy
  Efficient Machine Learning and Cognitive Computing}, 2019.

\bibitem{lan2020albert}
Z.~Lan \emph{et~al.}, ``{ALBERT}: A lite {BERT} for self-supervised learning of
  language representations,'' in \emph{Proc. ICLR}, 2020.

\bibitem{guo2020incorporating}
J.~Guo \emph{et~al.}, ``Incorporating {BERT} into parallel sequence decoding
  with adapters,'' in \emph{Proc. NeurIPS}, 2020, pp. 10\,843--10\,854.

\bibitem{lee2022memory}
J.~Lee \emph{et~al.}, ``Memory-efficient training of {RNN-Transducer} with
  sampled softmax,'' in \emph{Proc. Interspeech}, 2022, pp. 4441--4445.

\bibitem{sainath2019two}
T.~N. Sainath \emph{et~al.}, ``Two-pass end-to-end speech recognition,'' in
  \emph{Proc. Interspeech}, 2019, pp. 2773--2777.

\bibitem{wang2021streaming}
T.~Wang \emph{et~al.}, ``Streaming end-to-end {ASR} based on blockwise
  non-autoregressive models,'' in \emph{Proc. Interspeech}, 2021, pp.
  3755--3759.

\bibitem{touvron2023llama}
H.~Touvron \emph{et~al.}, ``{LLaMA}: Open and efficient foundation language
  models,'' \emph{arXiv preprint arXiv:2302.13971}, 2023.

\end{thebibliography}

\end{document}